\documentclass[11pt]{article}
\usepackage{graphicx} 
\usepackage[english]{babel}
\usepackage{amsmath}
\usepackage{amstext}
\usepackage{soul}
\usepackage{graphicx}%
\usepackage{amsfonts}%
\usepackage{amssymb}
\usepackage{dsfont}
\usepackage{rotating}
\usepackage{xy}
\usepackage{pdflscape}
\usepackage{url}
\usepackage[round, authoryear]{natbib}
\usepackage{mathrsfs}
\usepackage[colorlinks,citecolor=blue,urlcolor=blue]{hyperref}
\usepackage{setspace}
\usepackage{bm}
\usepackage{booktabs}
\usepackage{subcaption}
\usepackage{multirow}
\usepackage[margin=1in]{geometry}

\usepackage[pagewise]{lineno}

\newcommand{\keywords}[1]{\par\addvspace\baselineskip
                          \noindent{\bf Keywords:}\enspace\ignorespaces#1}

\newcommand{\bw}{\bm{w}}
\newcommand{\bW}{\bm{W}}
\newcommand{\bx}{\bm{x}}
\newcommand{\bX}{\bm{X}}

\newcommand{\Bezier}{B\'ezier }

\newcommand{\btheta}{\bm{\theta} }

\newcommand{\given}{\,|\,}
\newcommand{\trans}{^\text{T}}

\newcommand{\iid}{\stackrel{\text{iid}}{\sim}}

\title{Semiparametric Estimation of the Shape of the Limiting Bivariate Point Cloud}
\author{Reetam Majumder \\
SECASC \\
North Carolina State University \\
\and 
Benjamin A. Shaby \\
Department of Statistics \\
Colorado State University \\
\and Brian J. Reich \\
Department of Statistics \\
North Carolina State University \\
\and Daniel S. Cooley \\
Department of Statistics \\
Colorado State University
}
\date{}

\begin{document}
\maketitle
	
\begin{abstract}
We propose a model to flexibly estimate joint tail properties by exploiting the convergence of an appropriately scaled point cloud onto a compact limit set.  Characteristics of the shape of the limit set correspond to key tail dependence properties.  We directly model the shape of the limit set using \Bezier splines, which allow flexible and parsimonious specification of shapes in two dimensions.  We fit the \Bezier splines to data in pseudo-polar coordinates using Markov chain Monte Carlo sampling, utilizing a limiting approximation to the conditional likelihood of the radii given angles. We propose a novel prior on the shape of the limit set via constraints on the parameters of the \Bezier splines. A direct advantage of our Bayesian approach is that the support of this prior guarantees that each posterior sample is a valid limit set boundary, allowing direct posterior analysis of any quantity derived from the shape of the curve.  Furthermore, we obtain interpretable inference on the asymptotic dependence class by using mixture priors with point masses on the corner of the unit box.  Finally, we apply our model to bivariate datasets of extremes of variables related to fire risk and air pollution.
\end{abstract}

\section{Introduction}

  Multivariate tail risk calculations require knowledge of the strength of dependence in the joint tail of the relevant distribution.  Here, we propose a model to flexibly estimate joint tail characteristics in a way that coherently links several existing measures of tail dependence.  To do this, we describe tail dependence of a multivariate distribution through its associated \emph{gauge function} \citep{balkema-2010a,balkema-2010b,nolde-2014a}.  The homogeneity property of the gauge function allows us to recover the entire gauge function from its unit level set, which bounds the support of the appropriately scaled data points in the limit \citep{nolde-2022a,wadsworth-2022a}.  
We represent the unit level set of the gauge function using a semiparametric model specified within a Bayesian framework wherein required constraints on such functions are automatically satisfied. We obtain a posterior sample of gauge functions, and our prior specification ensures that each member of the sample is a valid gauge function not requiring any \emph{post hoc} adjustments such as re-scaling or truncation.

  Efforts to exploit the limit set representation of multivariate extreme values \citep{davis-1988a,kinoshita-1991a,balkema-2010a,balkema-2010b} have appeared only recently. \citet{wadsworth-2022a} decompose the data into pseudo-polar coordinates and use a limiting argument to approximate the distribution of the radii with a truncated gamma distribution whose parameters depend on the gauge function.  They first transform the data to unit exponential margins, and assuming a parametric form for the gauge function, perform maximum likelihood estimation with the truncated gamma likelihood.  They extend this approach using mixtures of parametric forms, but need to perform post-hoc re-scaling of the mixtures to satisfy the required properties of valid gauge functions.

  In contrast to \citet{wadsworth-2022a}, whose primary focus is estimating probabilities of sets in the joint tail region, \citet{simpson-2022a} focus on inference for the limit set boundary itself, taking a flexible semiparametric approach. They estimate the sample limit set by approximating the limiting upper endpoint of the distribution of radii with an estimated high quantile, as a function of angle. To do this, they fit a generalized Pareto distribution, whose scale parameter varies by angle, to the large radii. The radii are calculated by decomposing the bivariate data points transformed to unit exponential margins with a rank transformation. As the result is not a valid limit set, they perform a subsequent scaling and truncation procedure based on a Hill estimator \citep{Hill1975} to force their estimate to satisfy the required conditions.

  Like \citet{simpson-2022a}, our focus is flexible estimation of the limit set boundary, though our methodology is quite different. Here, we directly model the boundary of the limiting scaled point cloud, which is prescribed by the unit level set of the gauge function, as a \Bezier spline. \Bezier splines are constituted of \Bezier curves, with points on the curve that can be represented as Bernstein polynomials \citep[for reviews, see][]{hazewinkel-2012a,farouki-2012a}.  Similar semiparametric approaches have been used previously in multivariate extremes to characterize the Pickands dependence function for extremal dependence \citep{marcon2014, marcon2017, vettori2018}, the angular density in the context of multivariate regular variation \citep{Hanson2017}, and the angular dependence function \citep{Murphy-Barltrop2024}, which we will explore below as a direct byproduct of the gauge function. \Bezier splines are convenient here because they allow parsimonious specification of shapes in $\mathbb{R}^2$ which are defined by a small number of control points.  Placing appropriate constraints on the control points can ensure that the resultant shapes satisfy the conditions required of limit set boundaries.    To estimate the parameters of the \Bezier spline, we use the result from \citet{wadsworth-2022a} which says that, given a gauge function evaluated at the data points, the distribution of the large radial components decays like a gamma distribution whose rate parameter depends on the gauge function.  We then use standard Markov chain Monte Carlo machinery to sample from the posterior distribution.

  In this work, we place prior constraints on parameters of the \Bezier splines that define the limit set boundary, which induces a prior on the shape of the limit set. This work is thus far unique in that the support of this prior guarantees that every element of the posterior satisfies the challenging restrictions on the shape of the limit set \citep{nolde-2014a}.  Our approach has several advantages.  First, since we model the shape of the limiting point cloud in a way that automatically results in a valid limit set, we can use standard MCMC methods to obtain valid posterior inference.
  Second, our model allows the boundary of the limit set to exactly touch the corners of the unit box; this in particular gives a clean interpretation of the distinction between asymptotic independence (AI) and asymptotic dependence (AD) classes, since this distinction essentially corresponds to whether or not the boundary touches the upper right corner. Third, our approach produces a posterior sample of valid limit set curves which yields a realistic picture of the state of knowledge about the joint tail region given the data. Bayesian hypothesis testing can then be used  to study the properties of the limit set. In addition, we also note that our work builds on a growing literature on Bayesian approaches to multivariate extreme value analysis \citep[e.g.][to name a few]{BoldiDavison2007,sabourin-2013a, sabourin-2014a,deCarvalho2022,PadoanRizzelli2022}.

  The rest of the paper is arranged as follows. Section 2 introduces the limiting scaled point cloud and how it can be used to model the tail behavior of bivariate data. Section 3 develops the modeling of the limit set boundary using \Bezier splines. Section 4 contains a simulation study demonstrating our approach. Section 5 contains two applications---the Santa Ana Winds dataset, and ozone concentration data for the contiguous US---where we use \Bezier splines to model the tail dependence in the data. Section 6 concludes.

\section{The Limiting Scaled Point Cloud}\label{s:background}

Consider a collection of $n$ independent random vectors in $\mathbb{R}^2_+$, $\bX_1, \ldots, \bX_n$, each having joint density $f_{\bX}$, with standard exponential margins. At times, it will be convenient to transform the components of $\bX = (X_1, X_2)\trans$ into pseudo-polar coordinates $(R, \bW)$, as $R = X_1 + X_2$, and $\bW  = \bX / R$. Note that for $\bW = (W_1,W_2), W_2 = 1-W_1$.

Now define the scaled point cloud as the collection of points divided by $\log{n}$, $\{\bX_1 / \log{n}, \ldots, \bX_n/\log{n}\}$.  If we assume that $\lim_{t \rightarrow \infty} -\log f_{\bX}(t\bx) / t = g(\bx)$, $\bx \in \mathbb{R}^2_+$, for some continuous function $g$, then the scaled point cloud converges onto a compact limit set
\[
  G = \{\bx \in \mathbb{R}^2 : g(\bx) \leq 1\}
\]
as $n \rightarrow \infty$ \citep{davis-1988a,kinoshita-1991a,balkema-2010b,nolde-2014a,nolde-2022a}.  The function $g$ is called the \emph{gauge function} associated with the density $f_{\bX}$.  Denote the boundary of $G$ as $\partial G$.

Every gauge function $g$ is homogeneous of order one, with $g(c\bx) = cg(\bx)$ for any $c > 0$  \citep{nolde-2014a}.  We will exploit this property by modeling the limit set boundary $\partial G$ directly and using its associated gauge function, induced by homogeneity, for estimation (see Section \ref{sec:inference}).  Any valid limit set $G$ must satisfy the following constraints on its shape:
\begin{enumerate}
    \item \label{cond:starshaped} $G$ is \emph{star-shaped}, meaning that for any $t \in (0,1)$, if $\bx$ is in $G$, then $t\bx$ is also in $G$.
    \item \label{cond:touchbox} The supremum of the boundary $\partial G$ is 1 in each component direction.  That is, $\partial G$ touches, but does not cross, the upper and right-hand sides of the unit box.
\end{enumerate}

We seek a flexible way of representing the boundary $\partial G$ of the limit set $G$ that satisfies conditions \ref{cond:starshaped} and \ref{cond:touchbox} and can be estimated from \textit{iid} samples of the random vector $\bX$.
The shape of the limit set contains useful information about the extremal dependence of the distribution of the data.  \citet{nolde-2014a} linked particular features of the shape of $G$ with various indices of joint tail dependence in the literature.  The residual tail dependence coefficient \citep{ledford-1996a}, the angular dependence function \citep{wadsworth-2013a}, components of the conditional extremes model \citep{heffernan-2004a}, and the index $\tau_1(\delta)$ \citep{simpson-2020a} all have direct connections to the shape of $G$. Our primary focus is on the residual tail dependence coefficient, $\eta \in (0, 1]$, which is defined by assuming that, for $\bX$ in exponential margins, its survivor function satisfies
\[
  P(X_1 > x, X_2 > x) = \mathcal{L}(e^x)e^{-x/\eta}
\]
as $x \rightarrow \infty$, for some function $\mathcal{L}$ that is slowly varying at infinity \citep{ledford-1996a}.  Then the coefficient $\eta$ describes the strength of dependence in the joint tail, with $\eta \in (1/2, 1)$ indicating positive tail dependence but AI, and $\eta =1$ indicating AD, assuming $\mathcal{L}(x) \nrightarrow 0$.  The dependence class (AI vs. AD) is defined by the limiting conditional probability $\chi \in [0,1]$, where 
\[
  \chi := \lim_{x \rightarrow \infty} \frac{P(X_1 > x, X_2 > x)}{P(X_1 > x)},
\]
with $\chi=0$ characterizing AI, and $\chi > 0$ characterizing AD.

\begin{figure}
    \centering
    \includegraphics[width=0.5\textwidth,
    clip=true, trim=0 38 0 0]{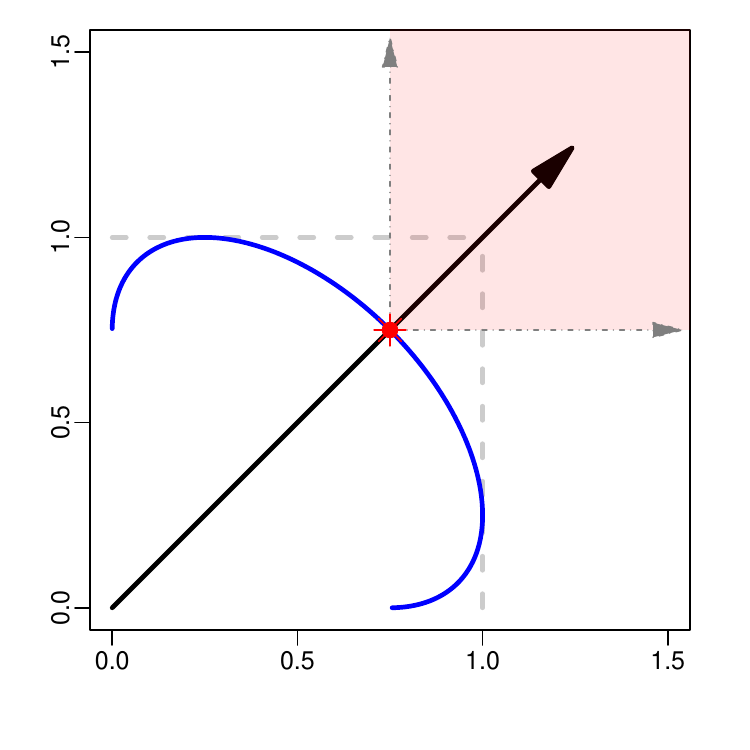}
    \caption{Schematic of $\eta$. The blue curve is the unit level set of gauge function $g(\bx)$, which forms the limit set boundary $\partial G$, and the proportional distance to the red point from the origin is the tail dependence coefficient $\eta$. While the red point is always on the diagonal, the intersection of the shaded red region and the blue curve does not necessarily occur on the diagonal.}\label{fig:eta-schematic}
\end{figure}

The residual tail dependence coefficient, $\eta$, can be calculated \citep{nolde-2014a, nolde-2022a} from shape of the limit set as 
\[
  \eta = \min\{r:r \times [1, \infty]^2 \cap G = \emptyset \}.
\]
  This is illustrated schematically in Figure \ref{fig:eta-schematic}, where one can think of sliding the shaded box down the ray with slope 1 until it first touches the boundary  $\partial G$. The radius corresponding to this first point of intersection is $\eta$.  Assuming as above that $\mathcal{L}(x) \nrightarrow 0$, when $\bX$ is asymptotically dependent, $\eta = 1$, so $\partial G$ necessarily touches the upper right-hand corner of the unit box.  Conversely, when $\bX$ is asymptotically independent, $\eta < 1$, so $\partial G$ does not touch the upper right-hand corner, and is referred to as \emph{blunt}.

\section{Modeling the Shape Using \Bezier Splines}\label{s:bezier}
\subsection{A \Bezier spline representation of the limit set boundary}\label{s:constraints}
\begin{figure}
    \centering
    \includegraphics[width=\textwidth]{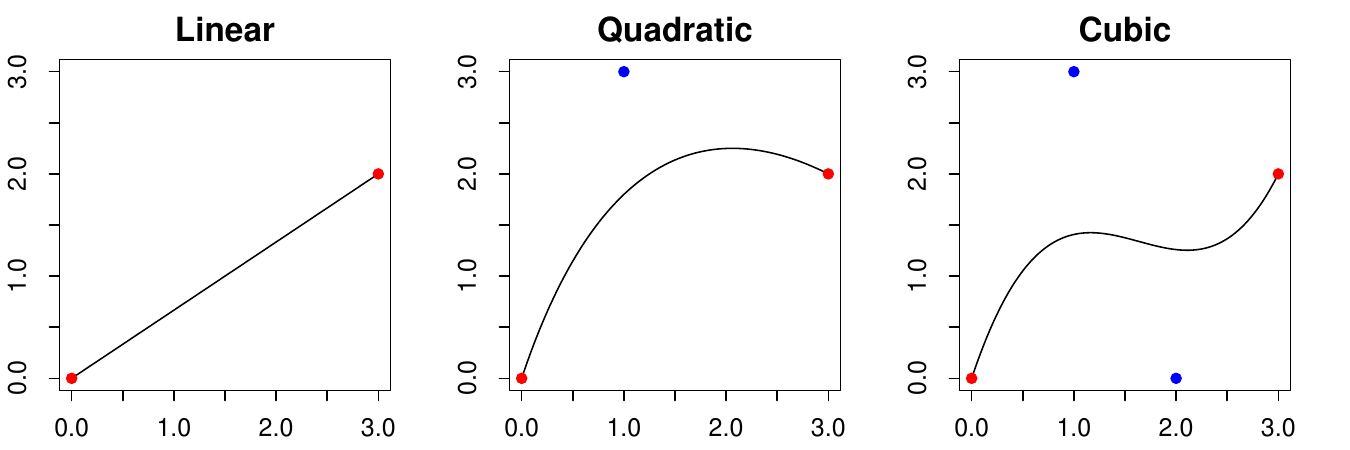}
    \caption{Examples of \Bezier curves of orders 1, 2, and 3.  The red control points (end points) $\mathbf{p}_0$ and $\mathbf{p}_m$ always lie on the curve, while the blue control points usually do not.}
    \label{f:bez_curves}
\end{figure}

\Bezier curves \citep[e.g.][]{hazewinkel-2012a,farouki-2012a} 
are a class of parametric functions that can be used as building blocks to represent complex shapes. \Bezier curves are defined by a set of \textit{control points} $\mathbf{p}_0$ to $\mathbf{p}_m$, where $m$ is the order of the curve. Figure \ref{f:bez_curves} plots examples of \Bezier curves of orders 1--3. The end points (red) define the beginning and end of the curve; intermediate control points (blue) of each curve  control its shape but generally do not lie on the curve. 
A quadratic \Bezier curve, for example, traces the path:
$$B(t) = (1-t)[(1-t)\mathbf{p}_0 + t\mathbf{p}_1] + t[(1-t)\mathbf{p}_1 + t\mathbf{p}_2],$$
for $0 \leq t \leq 1$. Rearranging this equation simplifies it to:
$$B(t) = (1-t)^2\mathbf{p}_0 + 2t(1-t)\mathbf{p}_1 + t^2\mathbf{p}_2.$$
A useful property is that if the three points are co-linear, then a quadratic \Bezier curve simplifies to a linear \Bezier curve.
Several \Bezier curves can in turn be linked together at the end points to form a \Bezier spline. The end points of each \Bezier curve within the spline now function as \textit{knots} for the spline. 
Splines comprised of quadratic \Bezier curves are particularly useful since analytical solutions for quadratic equations are straightforward to obtain, and therefore represent a flexible yet parsimonious way to approximate the shapes of limit set boundaries. In particular, increasing the order to cubic splines would make it difficult to constrain the shapes to the unit box, and would prevent the shapes from having the sharp corners required to represent asymptotically dependent limit set boundaries.

Because they are parsimoniously parameterized and straightforward to constrain, quadratic \Bezier splines are convenient for modeling the boundary $\partial G$ of the limit set $G$.  
We specify $\partial G$ as a \Bezier spline comprised of three quadratic \Bezier curves $g_B = \{B_1(t),B_2(t),B_3(t)\}$, where $B_1(t) := B(t;\mathbf{p}_0,\mathbf{p}_1,\mathbf{p}_2)$, $B_2(t) := B(t;\mathbf{p}_2,\mathbf{p}_3,\mathbf{p}_4)$, and $B_3(t):= B(t;\mathbf{p}_4,\mathbf{p}_5,\mathbf{p}_6)$, for $\mathbf{p}_i \in \mathbb{R}^2$, $i=0, 1, \ldots, 6$. The three curves trace the paths:
\begin{align*}
    B_1(t) = (1-t)^2\mathbf{p}_0 + 2t(1-t)\mathbf{p}_1 + t^2\mathbf{p}_2,\\
    B_2(t) = (1-t)^2\mathbf{p}_2 + 2t(1-t)\mathbf{p}_3 + t^2\mathbf{p}_4,\\
    B_3(t) = (1-t)^2\mathbf{p}_4 + 2t(1-t)\mathbf{p}_5 + t^2\mathbf{p}_6,
\end{align*}
for $0\leq t \leq 1$. We denote the point $\mathbf{p}_i := (p_{i,1},p_{i,2})$, $0\leq p_{i,1},p_{i,2} \leq 1$, and place two sets of constraints on the curves in order to elicit valid gauge functions which satisfy conditions \ref{cond:starshaped} and \ref{cond:touchbox}. The first set of constraints ensure that the \Bezier spline touches all four edges of the unit square:
\begin{align*}
    p_{0,1} &= p_{6,2} = 0,\\
    p_{2,2} &= p_{4,1} = 1.
\end{align*}
    The second set of constraints are sufficient conditions to ensure that the star-shaped property holds for the spline:
\begin{align*}
    p_{1,1} &\leq  p_{2,1},\\
    m(\mathbf{0},\mathbf{p}_1) &\geq m(\mathbf{0},\mathbf{p}_2),\\
    m(\mathbf{0},\mathbf{p}_4) &\geq m(\mathbf{0},\mathbf{p}_5),\\
    p_{4,2} &\geq p_{5,2},\\
    p_{3,1} &= p_{3,2},\\
    p_{3,1} &\geq \max(p_{2,1},p_{4,2})
\end{align*}

where $m(\mathbf{p},\mathbf{p'})$ denotes the slope of the line connecting the points $\mathbf{p}$ and $\mathbf{p'}$, and $\mathbf{0} = (0,0)$ is the origin. The final condition ensures that if $p_{2,1}$ or $p_{4,2}$ are 1, $p_{3,1}$ is also $1$. 
Thus, we arrive at a model for the limit set boundary $\partial G$, indexed by the 9 univariate parameters $\btheta_g = (p_{0,2}, p_{1,1}, p_{1,2}, p_{2,1}, p_{3,1}, p_{4,2}, p_{5,1}, p_{5,2}, p_{6,1})\trans$.
In practice, these constraints are enforced within a Metropolis sampler. If a candidate MCMC sample for one of the parameters does not satisfy the constraints, the sampler rejects it automatically.
\begin{figure}
    \centering
    \includegraphics[width=0.3\textwidth]{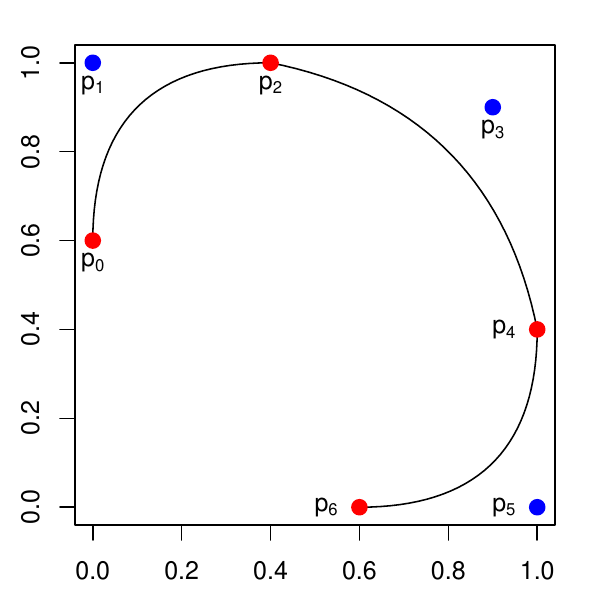}
    \includegraphics[width=0.3\textwidth]{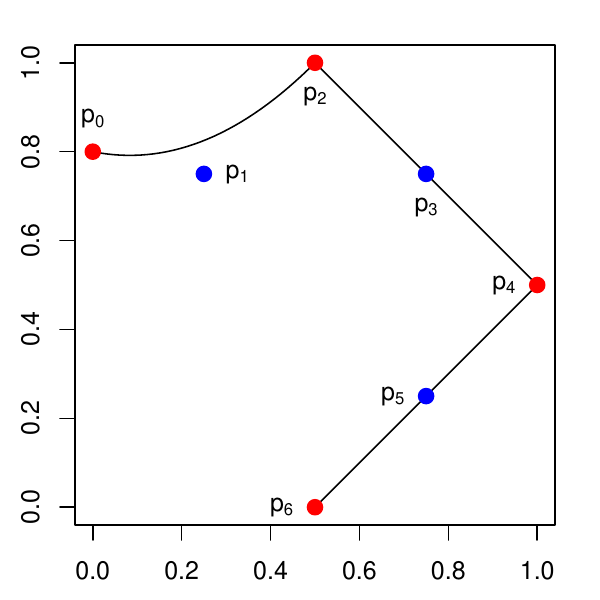}
    \includegraphics[width=0.3\textwidth]{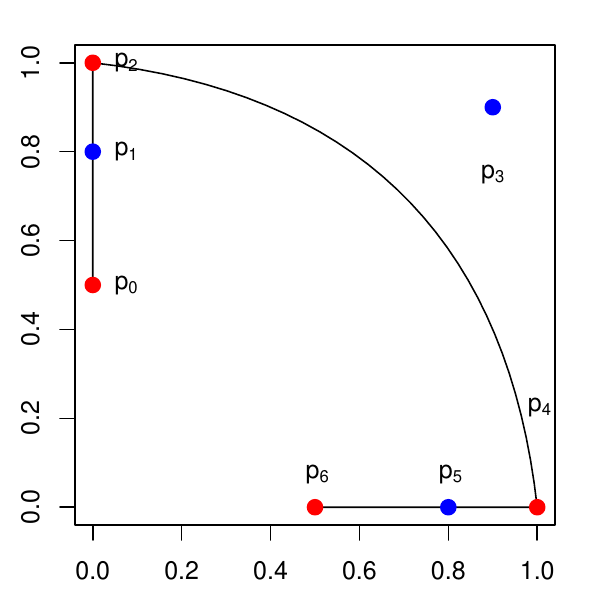}
    \includegraphics[width=0.3\textwidth]{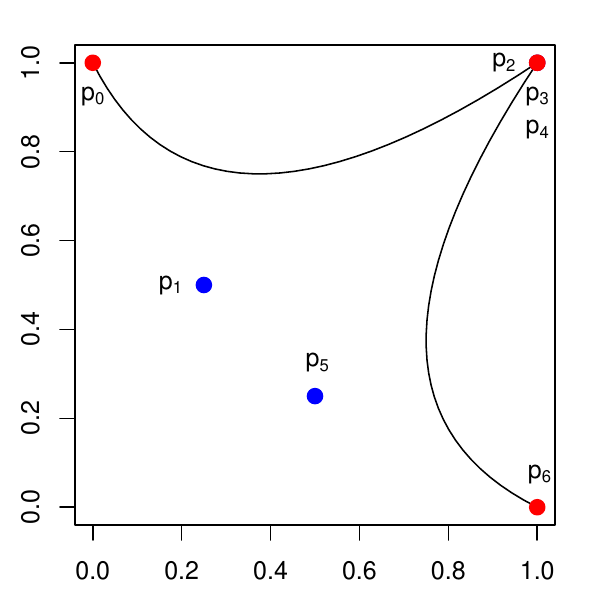}
    \includegraphics[width=0.3\textwidth]{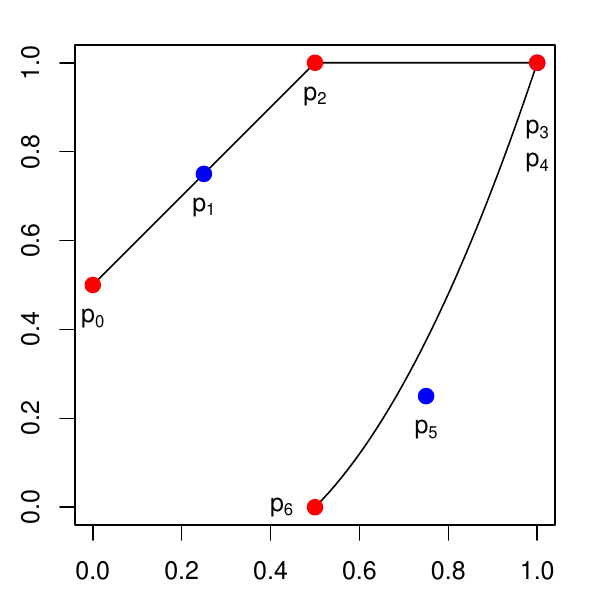}
    \includegraphics[width=0.3\textwidth]{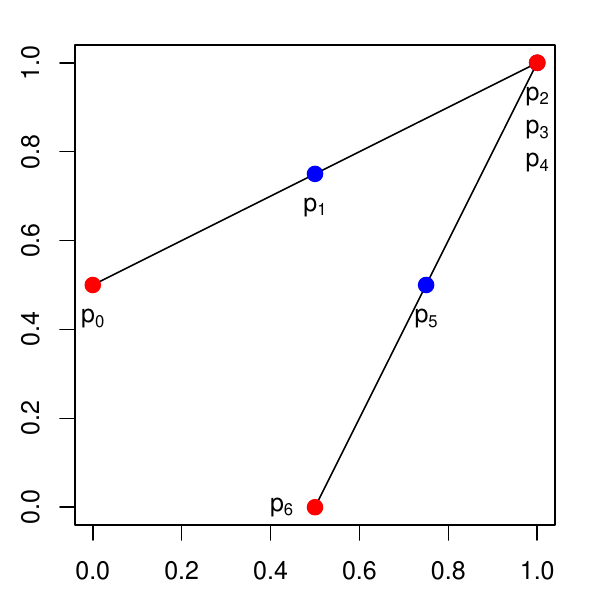}
    \caption{Examples of unit level sets of gauge functions that can be expressed using \Bezier splines comprised of 3 quadratic \Bezier curves. The red points are the end points of each curve, while the blue points are intermediate points controlling the shapes of the curves.}
    \label{f:bez_spline}
\end{figure}

Figure \ref{f:bez_spline} plots \Bezier splines under these constraints, each representing a gauge function with different dependence properties. Top row plots correspond to AI scenarios, whereas plots in the bottom row correspond to AD scenarios. The four red control points are the knots of the spline. The three blue control points affect the shape, and the spline passes through them only if they are co-linear with the preceding and proceeding control points. In the general case, there are 9 coordinates, each admitting a uniform support, which need to be estimated to fully specify a valid gauge function. Richer models can be achieved using more control points; however, it would come with increased computational cost and additional constraints to ensure that conditions \ref{cond:starshaped} and \ref{cond:touchbox} hold. The quadratic \Bezier spline with 4 knots therefore constitutes a parsimonious representation for $\partial G$ which is still flexible enough to capture multiple dependence regimes and mimic most of the common parametric models.

\subsection{Statistical inference for the limit set boundary}
\label{sec:inference}

With a model defined for the limit set boundary $\partial G$, we turn to the question of how to estimate the shape from \textit{iid} copies of the random vector $\bX$ in standard exponential margins.  After transforming $\bX$ to pseudo-polar coordinates $(R, \bW)$, a convenient form \citep{wadsworth-2022a} for the conditional density of a large radius $R$, given the angle $\bW$, is
\[
  f_{R \given W}(r \given \bw) \propto r^{d-1} \exp \{-rg(\bw)[1 + o(1)]  \}, \quad r \rightarrow \infty,
\]
where $d$ is the dimension of $\bX$ (we have only considered $d=2$ here). For likelihood-based inference, \citet{wadsworth-2022a} show that the $o(1)$ term can be moved outside the exponent in most cases and therefore ignored; they consequently consider the approximation adequate for radii larger than a threshold $r_0(\bw)$. This yields a truncated gamma likelihood:
\begin{equation}
  \label{eqn:trunc-gamma}
  R \given \bW = \bw, R > r_0(\bw), \btheta_g \sim \text{truncGamma}\left(\alpha, g_{\btheta_g}(\bw)\right).
\end{equation}
For most common bivariate copulas, $\alpha=2$. However, the quality of the approximation tends to vary \citep[for further details, see][]{wadsworth-2022a}, and $\alpha$ is usually treated as a parameter to be estimated. Thus, given a gauge function $g_{\btheta_g}$ an approximate likelihood for the large radii given the angles is
\begin{equation}
  \label{eqn:data-likelihood} 
  L(\btheta_g, \alpha; (r_1,\bw_1), \ldots, (r_{n_0},\bw_{n_0})) = 
    \prod_{i=1}^{n_0} \frac{g_{\btheta_g}(\bw)^\alpha}{\Gamma(\alpha)}
       \frac{r_i^{\alpha-1}\exp\{-r_ig_{\btheta_g}(\bw_i)\}}
            {1 - F(r_0(\bw_i); \alpha, g_{\btheta_g}(\bw_i))},
\end{equation}
where $n_0$ is the number of points exceeding the threshold $r_0(\bw)$, and $F(\,\cdot\,; \alpha, g_{\btheta_g}(\bw_i))$ is the CDF of a gamma distribution with shape parameter $\alpha$ and rate parameter $g_{\btheta_g}(\bw_i)$.

\begin{figure}
    \centering
    \includegraphics[width=0.5\textwidth,
    clip=true, trim=0 0 0 0]{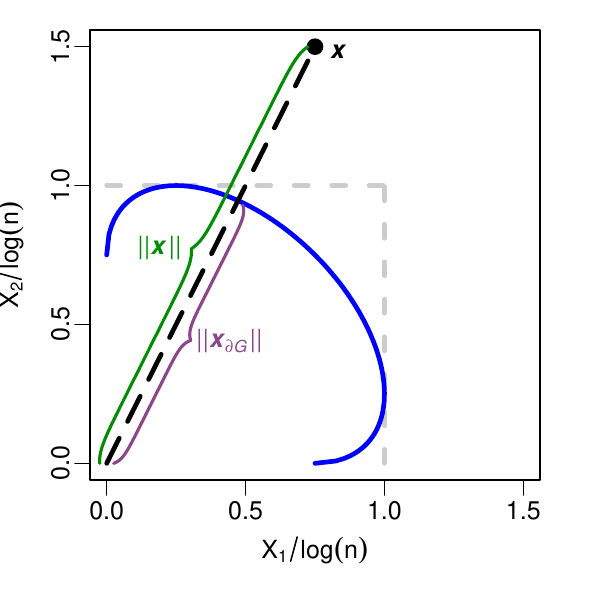}
    \caption{Schematic of how to calculate $g_{\btheta_G}$ at a data point $(\bx)$, given a boundary curve $\partial G$.  The value of the gauge function is the distance from the origin to $\bx$, relative to the distance from the origin of the intersection of the ray connecting $\bx$ with the origin and the boundary $\partial G$.}\label{fig:gauge-schematic}
\end{figure}

To calculate the gauge function at each data point, as required in the likelihood \eqref{eqn:data-likelihood}, we exploit the homogeneity property of $g$.  This gives us that the value of the gauge function evaluated at a point $\bx$ is the distance from the origin to $\bx$, relative to the distance from the origin of the intersection of the ray connecting $\bx$ with the origin and the boundary $\partial G$.  In the schematic in Figure \ref{fig:gauge-schematic}, the intersection with $\partial G$ is denoted as $\bx_{\partial G}$, so that 
\begin{equation}
  g_{\btheta_G}(\bx) = \frac{\| \bx \|}{\| \bx_{\partial G} \| }.
  \label{eqn:gauge-from-boundary}
\end{equation}

We also need to select a threshold $r_0(\bw)$, as a function of angle.  \citet{wadsworth-2022a} and \citet{simpson-2022a} both chose thresholds as functions of the angle, first as empirical quantiles of moving windows of angle and then using smooth semi parametric quantile regression.  We employ a much simpler approach, and choose a high quantile in each threshold marginal component. We have found that this very basic strategy results in estimation performance at least comparably as good as more complicated alternatives.  In addition, choosing marginal thresholds has two key advantages.  First, it is simple to implement and requires no intricate tuning.  Second, it permits, in principle, transformation to standard exponential margins within a hierarchical model, whereas thresholds that depend jointly on both components do not.  With this in mind, we choose a value $\tau \in (0,1)$, and then set each marginal threshold at the $\tau^\text{th}$ marginal empirical quantile $q_{\tau, X_1}$ for $X_1$ and $q_{\tau, X_2}$ for $X_2$.  In pseudo-polar coordinates, this gives a radial threshold of 
\[
  r_0(\bw) = \begin{cases}
    \frac{q_{\tau, X_2}}{1-w}, & \quad  w \in  \left[0, \frac{q_{\tau, X_1}}{q_{\tau, X_1} + q_{\tau, X_2}}\right] \\\\
    \frac{q_{\tau, X_1}}{w}, & \quad  w \in \left(\frac{q_{\tau, X_1}}{q_{\tau, X_1} + q_{\tau, X_2}}, 1\right].
  \end{cases}
\]
While our simple strategy for threshold construction could potentially fail in pathological cases where it spuriously ignores important observations, we have not encountered such cases. A sensitivity study comparing the effect of the choice of threshold on estimation of $\eta$ is provided in the Supplementary Material \citep[][Section A]{Bezier_supplement}.

\subsection{Prior distributions for control points}
Our model for the limit set boundary $\partial G$ is indexed by the 9 univariate parameters, viz. $\btheta_g = (p_{0,2}, p_{1,1}, p_{1,2}, p_{2,1}, p_{3,1}, p_{4,2}, p_{5,1}, p_{5,2}, p_{6,1})\trans$.
To inform prior selection for these control points, we examine the boundary $\partial G$ of four parametric copula models, and learn the conditions on the control points which allow the \Bezier curve to mimic their shapes. The four copulas that we consider are the Gaussian, inverted logistic, logistic, and asymmetric logistic. The first two are asymptotically independent, while the final two are asymptotically dependent; analytical expressions of dependence measures for these models are provided in Table \ref{t:copula}, replicated from \citet{simpson-2022a}.
Since the limit set boundary $\partial G$ can take on a variety of shapes, including the AD case where it touches the upper right-hand corner of the unit box \citep[see e.g.,][Figure 2]{nolde-2022a}, we let the coordinates of the control points (i.e., the members of $\btheta_g$) vary in $[0,1]$ (subject to the constraints presented in Section \ref{s:constraints}) for flexibility, which permits the possibility of them being exactly equal to 0 or 1 to accommodate AD as seen in the logistic and asymmetric logistic copulas and the very weak dependence as seen in the inverted logistic copula. We now outline the support for the distributions of the control points which will allow the \Bezier splines to mimic our four copulas of interest, and then specify priors for all parameters to be used in the remainder of this study.

The limit set boundary for an asymptotically dependent copula is obtained whenever $p_{2,1}=1$ or $p_{4,1}=1$; additionally, the logistic copula is implied by $\mathbf{p}_2 = \mathbf{p}_3 = \mathbf{p}_4 = (1,1)$ (Figure \ref{f:bez_spline}, bottom-right). This is equivalent to collapsing the second curve of the \Bezier spline to a single point, and can be incorporated into our model by having a semi-continuous prior distribution for $p_{2,1}, p_{3,1}, \mbox{ and } p_{4,2}$ with support over (0,1] which includes a point mass at 1. Similarly, collapsing the second curve and having a semi-continuous prior on $p_{0,2}$ and $p_{6,1}$ can incorporate an approximation of the limit set boundary for an asymmetric logistic copula. The theoretical shape of an asymmetric logistic copula is represented using a red line in Figure \ref{fig:gauge_euclidean} (bottom right panel), while \Bezier spline approximations can be seen in the form of the blue line in the same figure, as well as in  Figure \ref{f:bez_spline} (bottom left panel). In both cases, the \Bezier splines approximate the sharp angles in the interior using smooth quadratic curves.
Finally, approximating the limit set boundary  for an inverted logistic copula using a \Bezier spline would require the first and third curves to collapse onto the $x_1=0$ and $x_2=0$ lines respectively (Figure \ref{f:bez_spline}, top-right). To accommodate this case, we set semi-continuous priors with point masses at 0 for $p_{1,1}, p_{2,1},p_{4,2},\mbox{ and } p_{5,2}$. To flexibly accommodate the wide range of limit set boundary shapes possible within our framework, we set priors as follows:
\begin{align*}
    \alpha &\sim \mbox{LogNormal}(1,1),\\
    p_{0,2},p_{1,2}, p_{5,1},p_{6,1} &\iid \mbox{Uniform}(0,1).
\end{align*}
The LogNormal prior on $\alpha$ has its density concentrated near $\alpha = 2$. 
The remaining control points have priors that are the mixture of a standard Uniform distribution and at least one point mass (to allow important geometric features of $\partial G$ with positive probability). They have the following forms:
\begin{align*}
    p_{1,1} &\iid 0.1\cdot \mathbb{I}(p_{1,1}=0) + 0.9 \cdot \mbox{Uniform}(0,1), \\
    p_{2,1} &\iid 0.1\cdot \mathbb{I}(p_{2,1}=0) + 0.8\cdot \mbox{Uniform}(0,1) + 0.1\cdot \mathbb{I}(p_{2,1}=1),\\
    p_{3,1}\given p_{2,1},\, p_{4,2} &\sim 0.6\cdot \mbox{Uniform}(0,1) + 0.4\cdot \mathbb{I}(\max(p_{2,1},p_{4,2})=1).
\end{align*}
The points $p_{5,2}$ and $p_{4,2}$ are distributed identically to $p_{1,1}$ and $p_{2,1}$, respectively. The point mass probabilities were chosen on the basis of a sensitivity study whose aim was to have good discrimination between the asymptotically dependent and asymptotically independent cases based on posterior probabilities of $\eta = 1$, while also simultaneously being able to provide unbiased, consistent estimates of $\eta$ when $\eta<1$. In particular, the prior on $p_{3,1}$ ensures that it is 1 only if AD is implied by either $p_{2,1}$ or $p_{4,2}$. These prior assumptions can accommodate the logistic, inverted logistic, and asymmetric logistic copulas, as well as intermediate forms such as the Gaussian copula which do not require point masses.

\subsection{Additional bivariate extremal dependence measures}

Alongside the tail dependence coefficient $\eta$ \citep{ledford-1996a} discussed in Section \ref{s:background}, we consider two additional indices of tail dependence which can be derived from the gauge function. The first of these is the angular dependence function  $\lambda(\omega)$ which considers different scalings for the two components of $\bX$. In particular, \citet{wadsworth-2013a} considered asymptotic probabilities of the following form:
\[P(X_1>\omega x, X_2 > (1-\omega)x) = \mathcal{L}_{\omega}(e^x)e^{-x \lambda(\omega)},\]
for some function $\mathcal{L}_{\omega}$ that is slowly varying at infinity, $\omega\in [0,1]$, and $\lambda(\omega)\in (0,1]$ when $X_1$ and $X_2$ have positive dependence. The  function $\lambda(\omega)$ therefore captures both extremal dependence regimes, with AD implying the pointwise lower bound of $\lambda(\omega)$. Evaluation of $\lambda(\omega)$ for rays $\omega$ near 0 and 1 corresponds to regions where one variable is larger than the
other. In particular, $\lambda(\omega)$ is a generalization of $\eta$, with $\eta = 1/\{2 \lambda(1/2)\}$. \citet{Murphy-Barltrop2024} found that global estimators  (such as the Simpson-Tawn estimator) which simultaneously estimate $\lambda(\omega)$ for all values of $\omega$ tend to provide better estimates compared to pointwise estimators (such as the Hill estimator). The \Bezier spline estimator is also a global estimator of $\lambda(\omega)$, and examining the estimation of  $\lambda(\omega)$ is therefore a useful measure to compare it with the Simpson-Tawn estimator.

We also investigate the dependence measure $\tau_1(\delta)$ \citep{simpson-2020a}, given by:
$$P(X_1>x, X_2 \leq \delta x) = \mathcal{L}_\delta (e^x)e^{-x/\tau_1(\delta)},$$
for some function $\mathcal{L}_{\delta}$ that is slowly varying at infinity, with $\delta \in [0,1]$. $\tau_1(\delta)$ is monotonically increasing in $\delta$, with $\tau_1(1) = 1$. This dependence measure  characterizes the probability of $X_1$ being large while $X_2$ is of a smaller order. Specifically, if there exists a $\delta^*<1$ such that $\tau_1(\delta^*) = 1$, it implies that $X_1$ can be large while $X_2$ is small (with $\delta$ determining just \textit{how} small). If no such $\delta^*$ exists, then $X_1$ can be large only if $X_2$ is also large. We can define $\tau_2(\delta)$ analogously, and both $\tau_1(\delta),\tau_2(\delta) \in (0,1]$.  

Table \ref{t:copula} provides the analytical expressions for these measures for the four copulas that we consider in our study. Like with $\eta$, these dependence measures can be exactly deduced from limit set boundaries, and hence from gauge functions. Since the \Bezier splines are quadratic polynomials, we can easily calculate $\tau_1(\delta)$ and $\lambda(\omega)$ for any estimated limit set boundary, simply by finding the intersections of polynomials and lines, which have closed-form solutions. While we do not present results for $\tau_2(\delta)$ in our study, it can be calculated in the same manner as $\tau_1(\delta)$. We refer the reader to \citet{simpson-2022a} for a detailed discussion on how each of these measures can be obtained from gauge functions in more general settings.
\begin{table}
\small
    \begin{subtable}[h]{\textwidth}
        \centering
\begin{tabular}{cc}\toprule
Copula & $g(\bx) = g(x_1,x_2)$\\\midrule
Gaussian & $\{x_1+x_2-2\rho(x_1x_2)^{1/2}\}/(1-\rho^2)$ \\[4mm]
Logistic & $\gamma^{-1}\max(x_1,x_2) + (1-\gamma^{-1})\min(x_1,x_2)$ \\[4mm]
Inv-Logistic & $(x_1^{1/\gamma} + x_2^{1/\gamma})^{\gamma}$ \\[4mm]
Asy-Logistic & $\min \bigl(x_1+x_2, \gamma^{-1}\max(x_1,x_2) + (1-\gamma^{-1})\min(x_1,x_2)\bigr)$\\\bottomrule
\end{tabular}
       \caption{Gauge function $g$ for the four bivariate copulas.}
       \label{tab:copula_gauge}
    \end{subtable}
    \\
    \begin{subtable}{\textwidth}
        \centering
    \begin{tabular}{ccccc}\toprule
      Copula & $\eta$ &  $\lambda(\omega)$ & $\tau_1(\delta) = \tau_2(\delta)$ \\\midrule
         Gaussian & $(1+\rho)/2$ & $\begin{cases} 
       \max(\omega,1-\omega), &\mbox{ if } t_{\omega} \leq \rho^2\\
        \frac{1-2\rho\sqrt{\omega(1-\omega)}}{1-\rho^2}, &\mbox{ if } t_{\omega} \geq \rho^2
        \end{cases}$  &  $\begin{cases}
        1, &\mbox{ if } \delta \geq \rho^2\\
        \frac{1-\rho^2}{1+\delta-2\rho \sqrt{\delta}}, &\mbox{ if } \delta \leq \rho^2
        \end{cases}$\\[4mm]
         Logistic & 1 & $\max(\omega,1-\omega)$ & $(\gamma^{-1}+1-\gamma^{-1}\delta)^{-1}$ \\[4mm]
         Inv-Logistic & $2^{-\gamma}$& $\{ \omega^{1/\gamma} + (1-\omega)^{1/\gamma}\}^\gamma$&1\\[4mm]
         Asy-Logistic & 1 & $\max(\omega,1-\omega)$ & 1\\\bottomrule
    \end{tabular}
    \caption{Dependence measures for the four bivariate copulas. Here, $t_{\omega} = \min(\omega,1-\omega)/\max(\omega,1-\omega)$.}
    \label{t:copula_dep}
     \end{subtable}
     \caption{Gauge function $g$, and a summary of dependence measures for the four bivariate copulas used in our study. Table has been reproduced from \citet{simpson-2022a}.}
     \label{t:copula}
     \normalsize
\end{table}
\section{Simulation Study}
\subsection{Study Setup}
We demonstrate the appropriateness of using \Bezier splines to model the limit set boundary corresponding to the gauge function by means of a simulation study comparing its performance with the Simpson-Tawn estimator \citep{simpson-2022a}. 

We consider four bivariate copulas to generate data with exponential marginal distributions: the Gaussian, the logistic, the inverted logistic, and the asymmetric logistic.
The Gaussian copula is parameterized by its correlation $\rho \in [0,1)$, while the dependence parameter for the remaining three copulas is $\gamma \in (0,1)$.
Table \ref{t:copula} lists the gauge functions associated with each copula, as well as the set of corresponding extremal dependence coefficients $\{\eta,\lambda(\omega),\tau_1(\delta)\}$.  Further details for these four copulas can be found in \cite{simpson-2022a}. For each copula, we consider five parameter settings: $\rho, \gamma = \{0.3,0.4,0.5,0.6,0.7\}$. Smaller values of $\gamma$ correspond to stronger tail dependence, whereas larger values of $\rho$ lead to stronger tail dependence.

For each copula and parameter combination, we generate 100 datasets of $n=5,000$ data points each. The datasets are converted to pseudo-polar coordinates $(R,\bW)$, and the $n_0$ points that are above the $\tau=0.75$ quantile marginal threshold for at least one variable are used to model the gauge function for each dataset.
The $n_0$ radii are assumed to approximately distributed according to a truncated gamma distribution with a common shape parameter $\alpha$ and a rate parameter equal to an appropriate gauge function evaluated at the data point. We use Metropolis updates for all parameters and run 11,000 MCMC iterations for each dataset, discarding the first 1,000 as burn-in. All Metropolis updates are tuned to give an acceptance probability of 0.4, and posterior convergence is diagnosed based on the visual inspection of trace plots. We compare our estimated limit set boundaries with the Simpson-Tawn estimator in terms of how well they estimate the set of dependence coefficients outlined in Table~\ref{t:copula}. The Simpson-Tawn estimator for the datasets was evaluated using the default settings recommended by the authors in \citet{simpson-2022a}. We use the root mean square error (RMSE) to compare estimates of the scalar $\eta$, and the root mean integrated square error (RMISE) to compare estimates of the functions $\lambda(\omega)$ and $\tau_1(\delta)$. The methodology is implemented in \texttt{R} \citep{Rcran}; the code is available on GitHub through the \texttt{BezELS} package \citep{BezELS}.

\subsection{Parameter estimates}

\begin{figure}
\centering
\begin{subfigure}{0.9\textwidth}
\centering
     \includegraphics[width=0.33\textwidth]{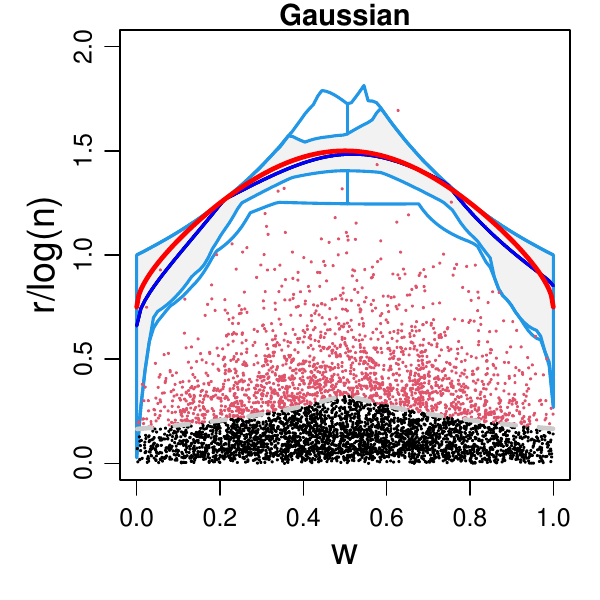}
     \includegraphics[width=0.33\textwidth]{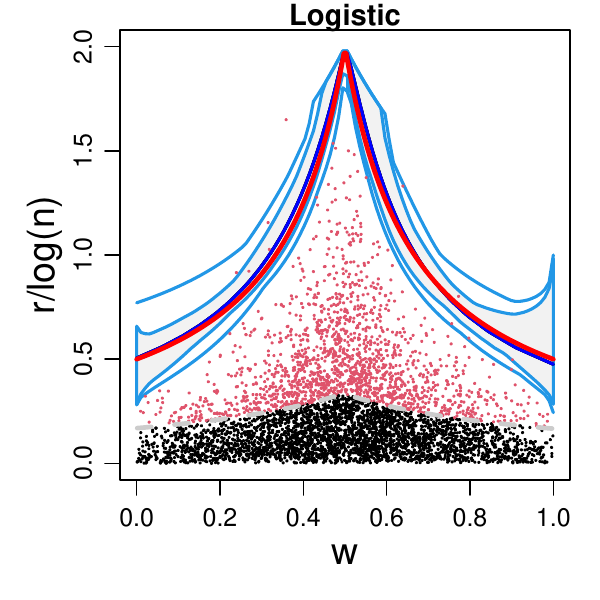}
     \\
     \includegraphics[width=0.33\textwidth]{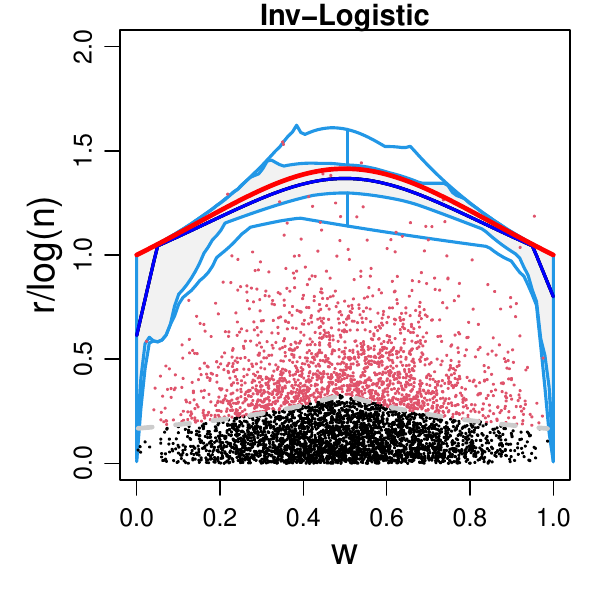}
     \includegraphics[width=0.33\textwidth]{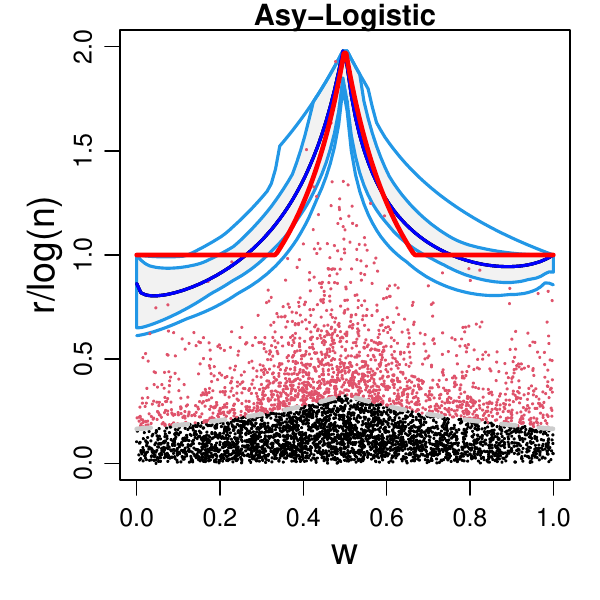}
    \caption{Estimated gauge functions in pseudo-polar space.}
    \label{fig:gauge_polar}
\end{subfigure}
\begin{subfigure}{0.9\textwidth}
\centering
     \includegraphics[width=0.33\textwidth]{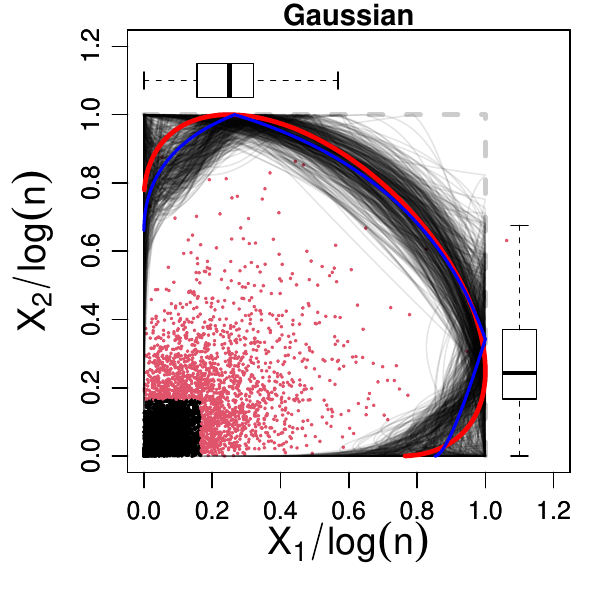}
    \includegraphics[width=0.33\textwidth]{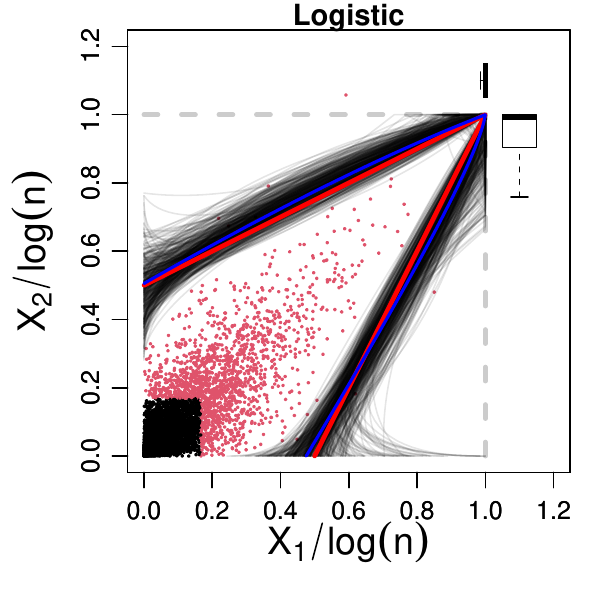}
    \\
    \includegraphics[width=0.33\textwidth]{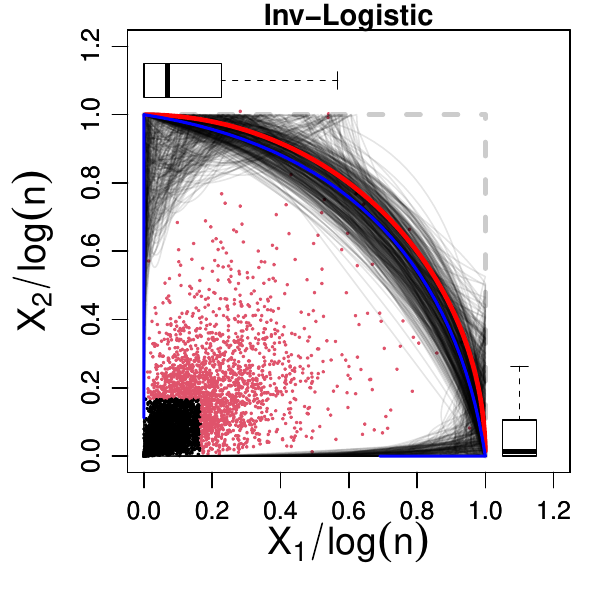}
    \includegraphics[width=0.33\textwidth]{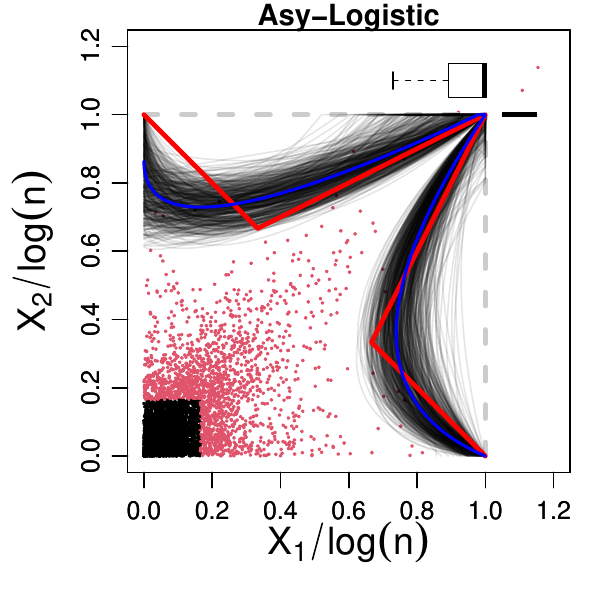}
    \caption{Estimated gauge functions in Euclidean space.}
    \label{fig:gauge_euclidean}
\end{subfigure}
\caption{Limit set boundaries based on \Bezier splines (blue) and corresponding functions for the data-generating model (red) in pseudo-polar and Euclidean space for the Gaussian, logistic, inverted logistic, and asymmetric logistic copulas with dependence parameters set to $0.5$}.
\label{fig:gauge_functions}
\end{figure}

Each panel in Figure \ref{fig:gauge_functions} plots the limit set boundaries elicited by the estimated \Bezier splines based on the posterior distribution from a single dataset with the dependence parameter set to 0.5. Plots in Figure \ref{fig:gauge_polar} display the dependence modeling in pseudo-polar coordinates. The dashed grey line corresponds to the threshold $r_0(\bw)$ for angles $\bw$. Each estimated limit set boundary  is represented as a functional boxplot \citep{HyndmanShang,SunGenton2011}, a visual representation of functional data analogous to a classical boxplot. Each functional quantile depicted in the figure is a function contained in the sample; in this case, the sample consists of limit set boundaries based on \Bezier splines, in pseudo-polar coordinates, drawn from the posterior. The curves are ordered according to a notion of modified band depth \citep{PintadoRomo2009}. The median is shown in dark blue, and the limit set boundary corresponding to the data-generating model is shown in red. The envelope represents the $50\%$ central region, analogous to the box of a classical boxplot. The outer light blue lines of the functional boxplot correspond to the whiskers of a classical boxplot. Finally, the vertical lines indicate the maximum envelope of the functions except outliers. Plots in Figure \ref{fig:gauge_euclidean} display the dependence models in Euclidean coordinates, with the median \Bezier spline in dark blue and the limit set boundary corresponding to the data-generating model in red. They are overlaid on \Bezier splines evaluated from 500 draws from the posterior distribution, plotted in gray.  The boxplots on the top and right margins correspond to the posterior distributions of $p_{2,1}$ and $p_{4,2}$ respectively, which serve as a visual indicator of asymptotic dependence in the data. Specifically, if the median of either boxplot is 1, the posterior median of $\eta$ is 1. The \Bezier splines are able to adequately represent the geometric form of all four copulas. 
\begin{figure}
    \centering
    \includegraphics[width=0.49\textwidth]{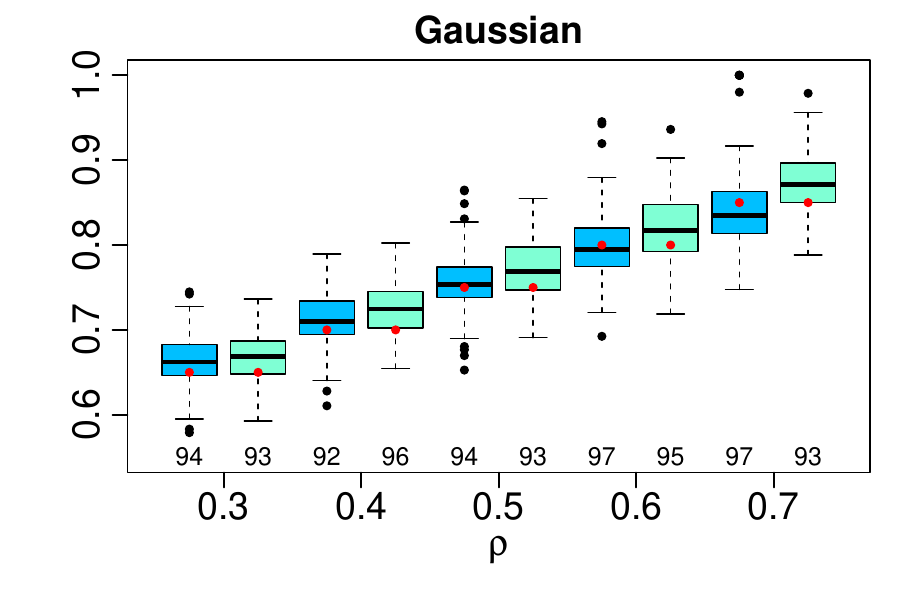}
    \includegraphics[width=0.49\textwidth]{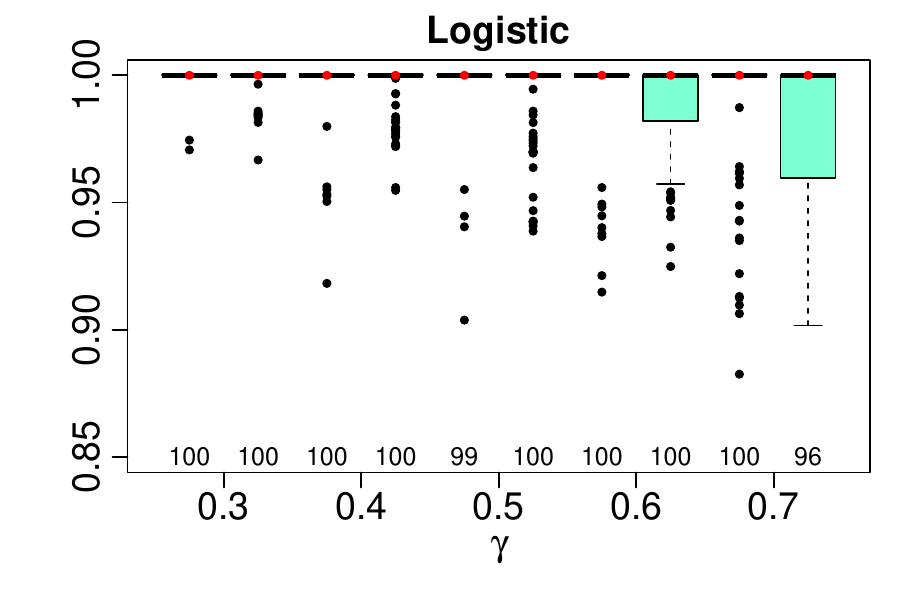}
    \includegraphics[width=0.49\textwidth]{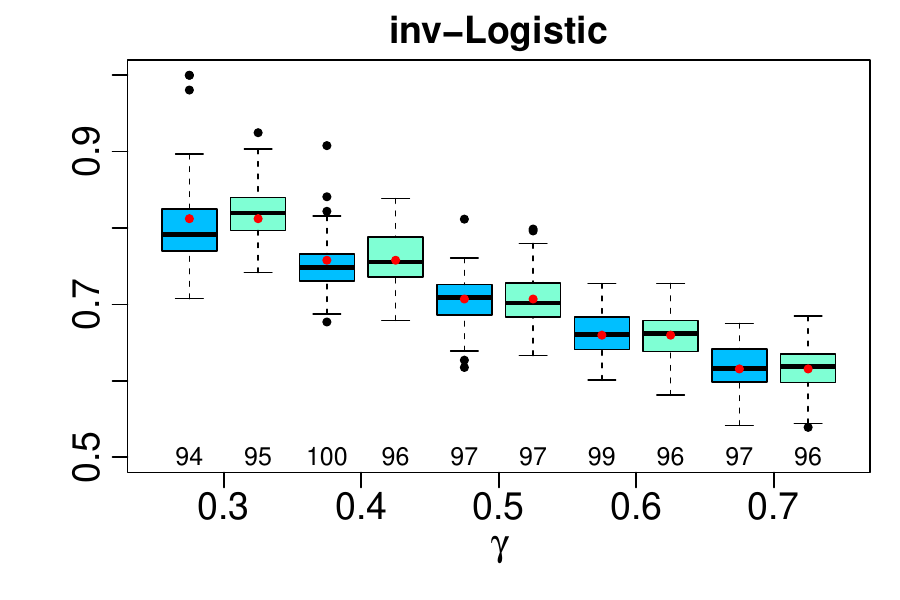}
    \includegraphics[width=0.49\textwidth]{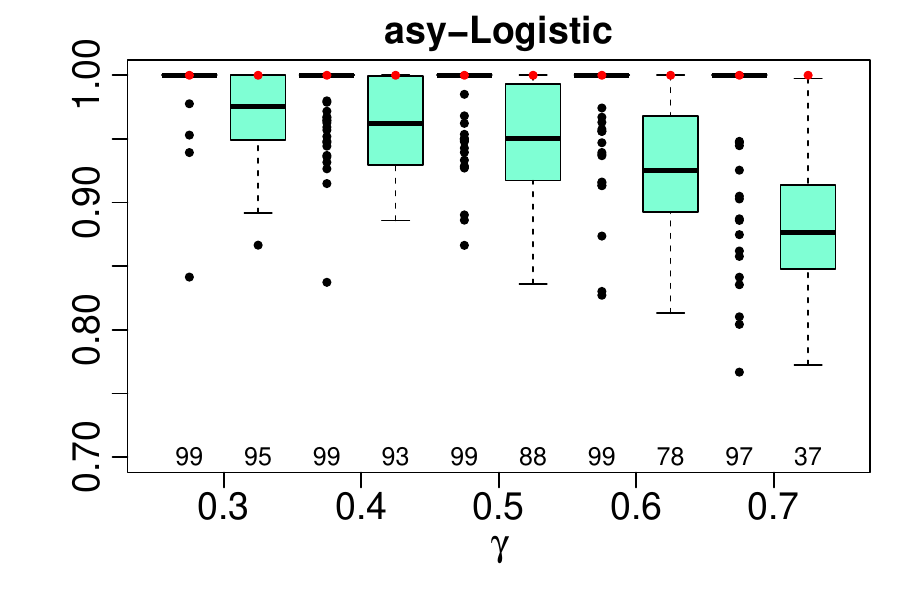}
    \caption{Sampling distribution of the posterior medians of $\eta$ based on the \Bezier spline estimate (blue) for the four copulas, alongside estimates using the Simpson-Tawn estimator (green). The red dots indicate the true values, and coverage of equi-tailed 95\% intervals are noted below each boxplot.}
    \label{f:MCMC_eta}
\end{figure}

Figure \ref{f:MCMC_eta} shows boxplots of the posterior median of $\eta$ for the four copulas based on the \Bezier spline (blue) and Simpson-Tawn estimators (green). Analytical values of $\eta$ obtained using expressions in Table \ref{t:copula} are shown as red dots in each plot, and the coverage of equi-tailed 95\% intervals are noted in plain text below each boxplot. Coverage for the Simpson-Tawn estimator is based on 100 bootstrapped samples from each of the 100 datasets. We plot the median instead of the mean for the \Bezier spline estimates since the posterior distributions are often highly asymmetric due to point-mass prior distributions. The \Bezier spline estimator has low bias and nominal coverage for estimates of $\eta$. The Simpson-Tawn estimator has nominal or near-nominal coverage in most cases except for the asymmetric logistic copula when the dependence parameter $\gamma$ is 0.5 or higher. It also has noticeably higher bias than the \Bezier spline estimator for both asymptotically dependent copulas, and shows a sharp decline in coverage as the strength of dependence drops for the asymmetric logistic copula.
\begin{figure}
    \centering
    \begin{subfigure}{0.9\textwidth}
    \includegraphics[width=0.49\textwidth]{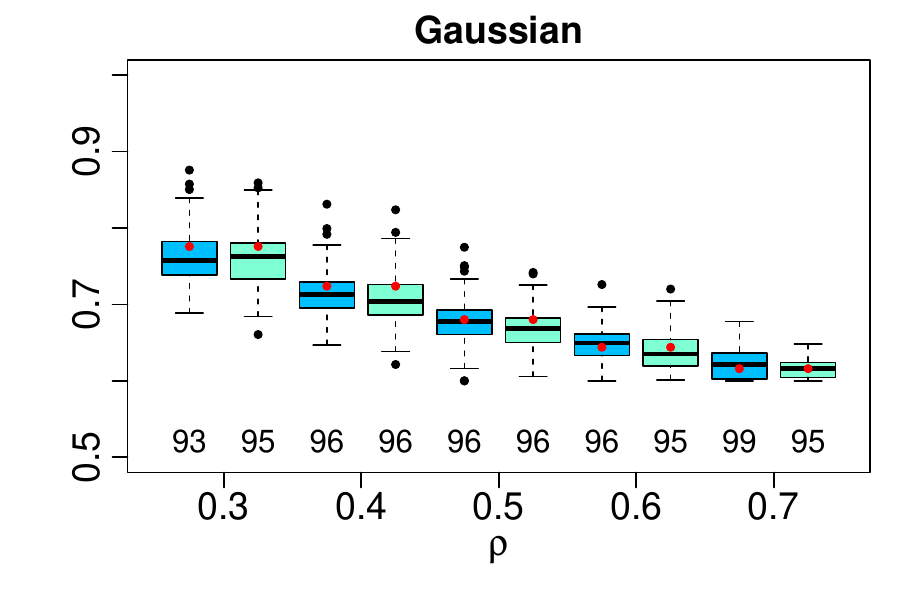}
    \includegraphics[width=0.49\textwidth]{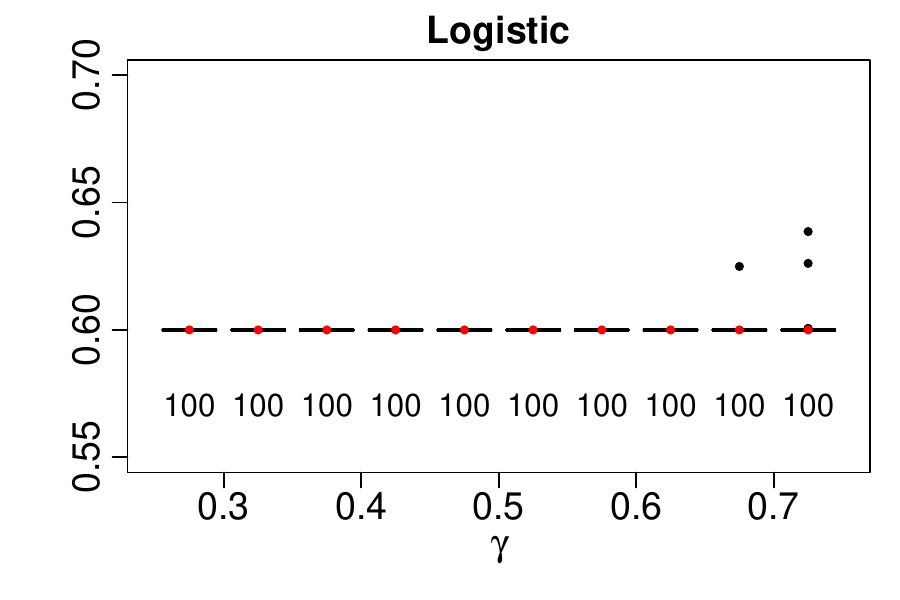}
    \includegraphics[width=0.49\textwidth]{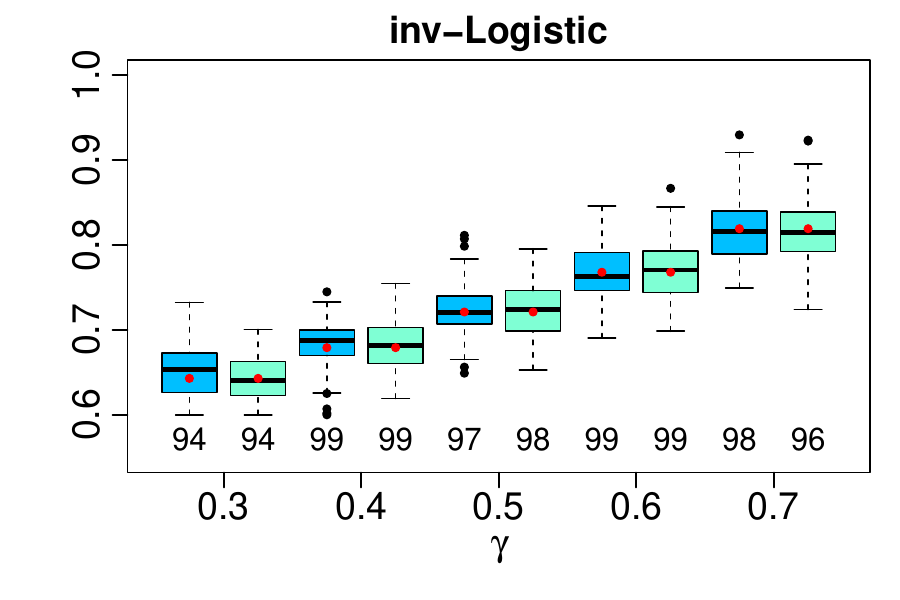}
    \includegraphics[width=0.49\textwidth]{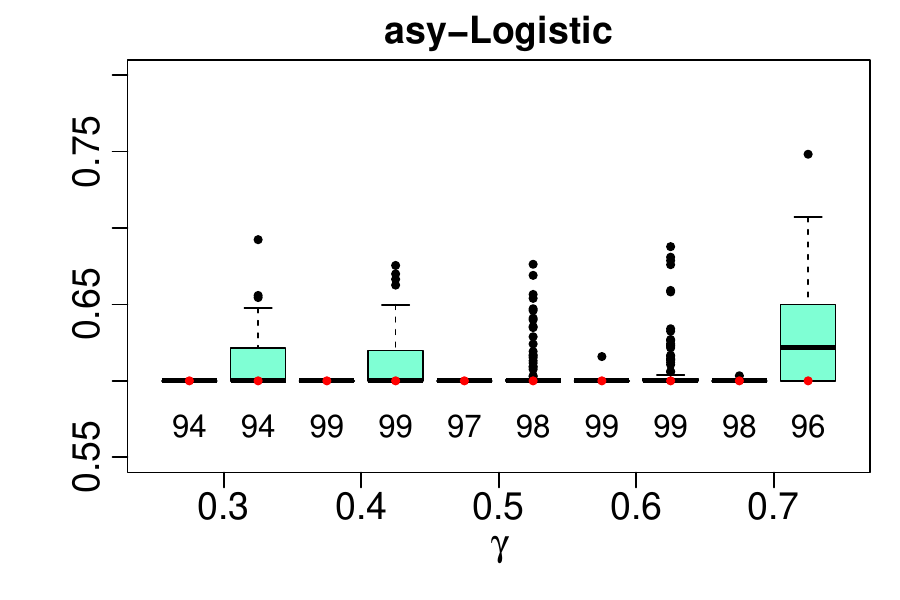}
    \caption{Sampling distribution of the posterior medians of $\lambda(0.40)$.}
    \label{f:MCMC_lambda}
    \end{subfigure}
\begin{subfigure}{0.9\textwidth}
    \includegraphics[width=0.49\textwidth]{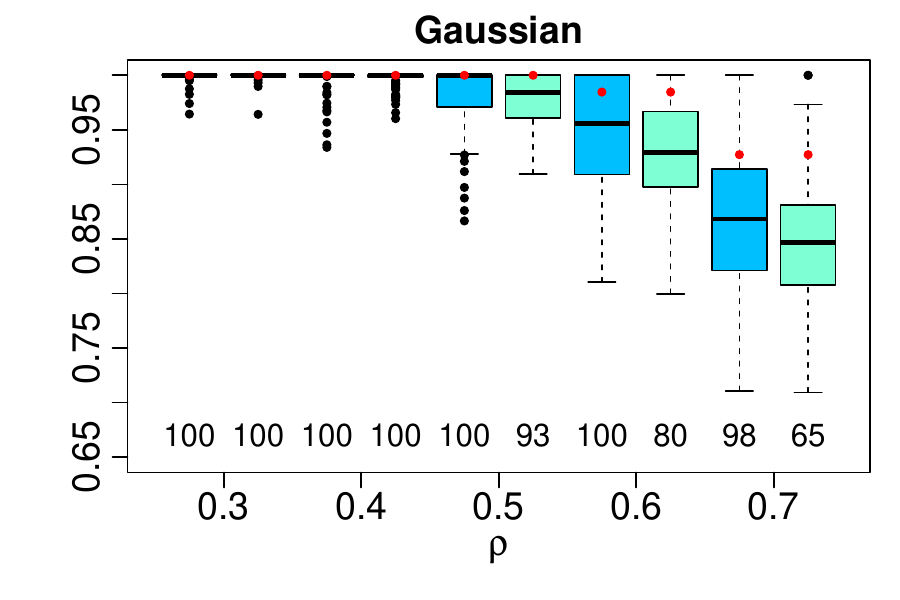}
    \includegraphics[width=0.49\textwidth]{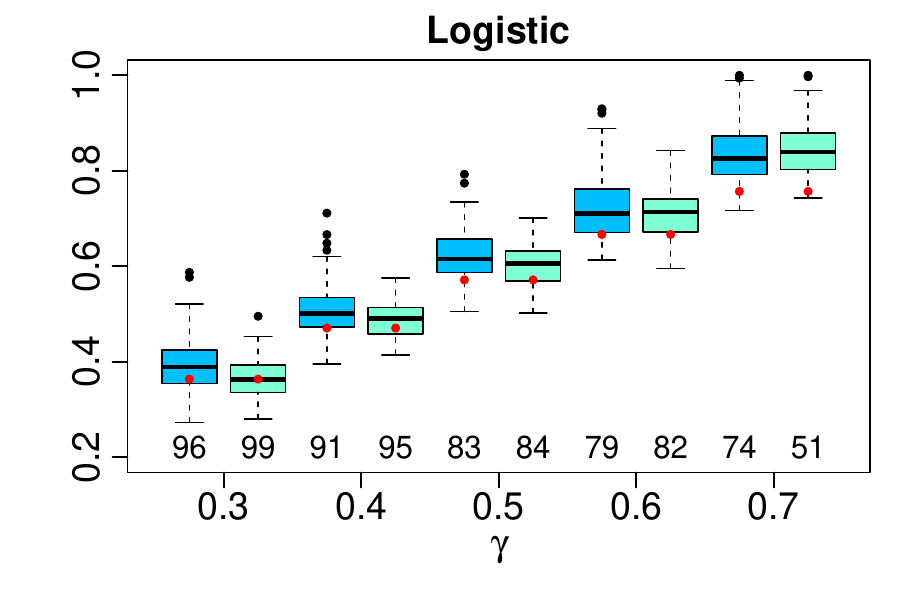}
    \includegraphics[width=0.49\textwidth]{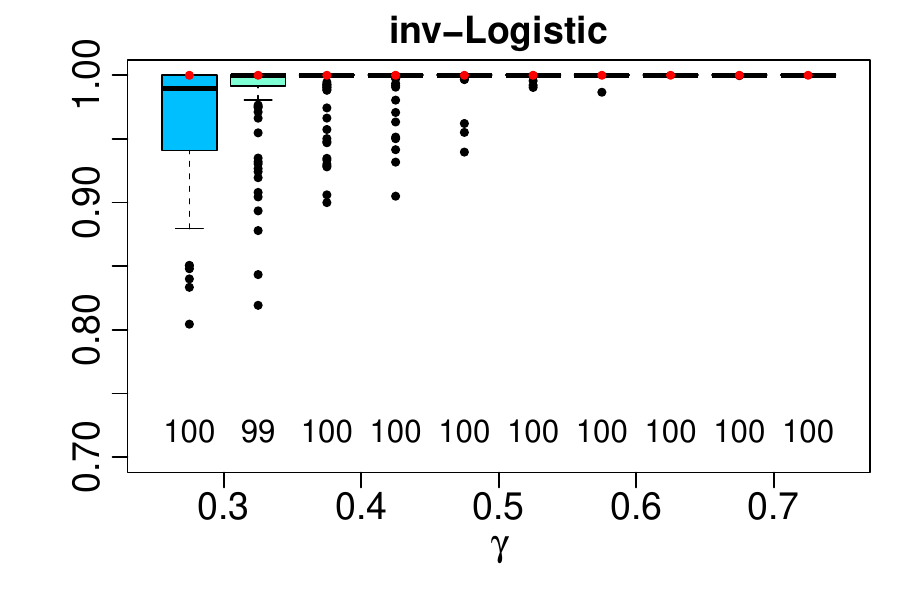}
    \includegraphics[width=0.49\textwidth]{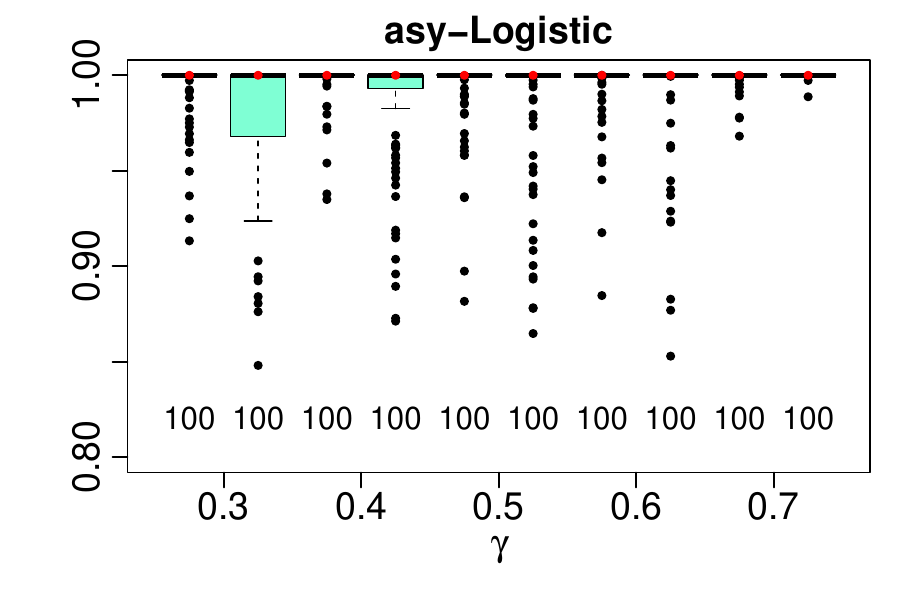}
    \caption{Sampling distribution of the posterior medians of $\tau_1(0.25)$.}
    \label{f:MCMC_tau}
    \end{subfigure}
     \caption{Sampling distribution of the posterior medians of $\lambda(0.40)$ and $\tau_1(0.25)$ based on the \Bezier spline (blue) and Simpson-Tawn (green) estimators for four copulas, with dependence parameters set to 0.5. The red lines indicate the true values, and coverage of equi-tailed 95\% intervals are noted below each boxplot.}
    \label{f:MCMC_lambda_and_tau}
\end{figure}
We evaluate $\lambda(\omega)$ and $\tau_1(\delta)$ based on the \Bezier spline and Simpson-Tawn estimators for $\omega,\delta = 0.01,0.02,\ldots,0.99$. Figure \ref{f:MCMC_lambda} shows boxplots for the posterior medians of $\lambda(0.40)$ based on the two estimators, with the corresponding dependence parameter set to 0.5. Figure \ref{f:MCMC_tau} similarly plots the distribution of $\tau_1(0.25)$ for the two estimators. Both estimators are better at estimating $\lambda(\omega)$ than $\tau_1(\delta)$, and the \Bezier spline estimator tends to have better coverage than the Simpson-Tawn estimator.

Table \ref{t:rmise} summarizes the RMSE ratio for estimates of $\eta$ and RMISE ratios for estimates of $\lambda(\omega)$ and $\tau_1(\delta)$ based on the \Bezier spline and Simpson-Tawn estimators. Most of the values are greater than 1, indicating that dependence measures based on \Bezier spline estimates of the gauge function have comparable or better RMSE/RMISE than those based on the Simpson-Tawn estimator. However, for the inverted logistic copula, the Simpson-Tawn estimator outperforms the \Bezier spline estimator when estimating of $\tau_1(\delta)$. Our experiments indicate that the error arises when the posterior $p_{4,2} > \delta$; this leads to $\tau_1(\delta)$ estimates to be less than 1, whereas the theoretical value for the inverted logistic copula is 1 for all $\delta$. Our approach, however, is still able to correctly estimate $\tau(\delta) = 1$ in all the inverted logistic scenarios for all but extremely small values of $\delta$.

\begin{table}
\centering
\small
\caption{RMSE ratios for estimates of $\eta$ and RMISE ratios for estimates of $\lambda(\omega)$ and $\tau_1(\delta)$ based on the \Bezier spline $(\hat{\eta},\hat{\lambda},\mbox{ and }\hat{\tau})$ and Simpson-Tawn $(\tilde{\eta},\tilde{\lambda},\mbox{ and }\tilde{\tau})$ estimators over simulated datasets for four copulas and five dependence levels.}
\label{t:rmise}
\begin{tabular}{ccccccc}\toprule
\textbf{Measure} & \textbf{Copula} & \multicolumn{5}{c}{\textbf{Dependence parameter value}} \\
 &  & 0.3 & 0.4 & 0.5 & 0.6 & 0.7 \\\midrule
\multirow{4}{*}{$\frac{RMSE(\tilde{\eta})}{RMSE(\hat{\eta})}$} & Gaussian & 1.02 & 1.11 & 1.04 & 1.06 & 0.81 \\
 & Logistic & 1.38 & 0.81 & 1.36 & 1.17 & 1.07 \\
 & Inv-Logistic & 0.70 & 0.99 & 1.11 & 1.03 & 1.01 \\
 & Asy-Logistic & 2.39 & 2.01 & 2.23 & 2.37 & 2.39 \\\midrule
\multirow{4}{*}{$\frac{RMISE(\tilde{\lambda})}{RMISE(\hat{\lambda})}$} & Gaussian & 1.19 & 1.18 & 1.05 & 0.94 & 0.75 \\
 & Logistic & 1.35 & 0.61 & 1.10 & 1.04 & 1.28 \\
 & Inv-Logistic & 0.82 & 1.10 & 1.20 & 1.10 & 1.04 \\
 & Asy-Logistic & 6.17 & 6.65 & 10.91 & 7.43 & 5.67 \\\midrule
\multirow{4}{*}{$\frac{RMISE(\tilde{\tau})}{RMISE(\hat{\tau})}$} & Gaussian & 0.85 & 1.00 & 1.08 & 1.17 & 1.18 \\
 & Logistic & 1.00 & 0.95 & 0.89 & 0.83 & 1.00 \\
 & Inv-Logistic & 0.23 & 0.27 & 0.17 & 0.28 & 0.17 \\
 & Asy-Logistic & 2.38 & 2.96 & 1.77 & 1.72 & 1.23\\\bottomrule
\end{tabular}
\normalsize
\end{table}

\begin{table}
\small
\centering
\caption{Number of datasets (out of 100) where the posterior median of $\eta$ is 1 for each scenario. Values in parenthesis correspond to the Simpson-Tawn estimator.}
\label{t:eta_estimates}
\begin{tabular}{cccccc}\toprule
 & \multicolumn{5}{c}{\textbf{Dependence parameter value}} \\
 & 0.3 & 0.4 & 0.5 & 0.6 & 0.7 \\\midrule
Gaussian & 00 (00) & 00 (00) & 00 (00) & 00 (00) & 06 (00) \\
Logistic & 98 (91) & 93 (79) & 96 (77) & 91 (74) & 82 (63) \\
Inv-Logistic & 02 (00) & 00 (00) & 00 (00) & 00 (00) & 00 (00) \\
Asy-Logistic & 96 (30) & 80 (25) & 82 (22) & 83 (14) & 84 (00)\\\bottomrule
\end{tabular}
\normalsize
\end{table}

Table \ref{t:eta_estimates} provides the number of datasets (out of 100) in each scenario where the posterior median of $\eta$ is estimated to be 1. The values in the parentheses are corresponding point estimates of $\eta$ based on the Simpson-Tawn estimator. For the \Bezier spline estimators, the values were always estimated to be near 0 for the asymptotically independent copulas, and always high for the asymptotically dependent copulas. While there were some cases where the \Bezier spline  estimates the posterior of $\eta$ as 1 when the dependence is high in an asymptotically independent copula, both methods are good at estimating AI correctly. On the other hand, the Simpson-Tawn estimates show a decline in their ability to estimate the correct value of $\eta=1$ for the asymptotically dependent copulas when dependence is low. This is especially noticeable for the asymmetric logistic copula, and has been documented by \citet{simpson-2022a} as well. The \Bezier spline estimator is much better at predicting AD correctly across all scenarios. We conclude that the \Bezier splines are adept at representing limit set boundaries associated with common parametric copula models, and are also flexible enough to represent a wider variety of edge cases. In all cases, the true value of $\eta$ was well-estimated from the posterior distribution, and the model is particularly adept at identifying AD.

\subsection{Additional simulation studies}
Two additional simulation studies are presented in the Supplementary Material \citep{Bezier_supplement}. In both cases, the results are compared based on RMSE/RMISE values of bivariate extremal dependence coefficients, as well as how often it estimates $\eta = 1$. The first study considers how our relatively straightforward method of selecting the quantile threshold affects the estimation process. This is carried out by comparing it against an `oracle' threshold which is an asymptotic approximation to the true conditional quantile $q_\tau(w)$ and requires knowledge of the true gauge function to compute. We are unable to find any meaningful improvement in the estimates when we consider an oracle threshold, which indicates that our choice of threshold is adequate for the scenarios we have considered. The second study repeats the simulation study presented in this section, but for a sample size $n=600$. This is carried out to ensure that our approach is still valid for small data sizes like the one that arises in our second application, presented in the following section. Our results indicate that despite slightly higher RMISE values and slightly lower coverage, the \Bezier splines can still capture the shape of the true limit set boundary with low bias. Comparisons with the Simpson-Tawn estimator provide results that are quite similar to the ones presented in this section.

\section{Applications}\label{s:application}

\subsection{Analysis of the Santa Ana Winds data}\label{s:SantaAna}

\begin{figure}
\centering
     \includegraphics[width=0.49\textwidth]{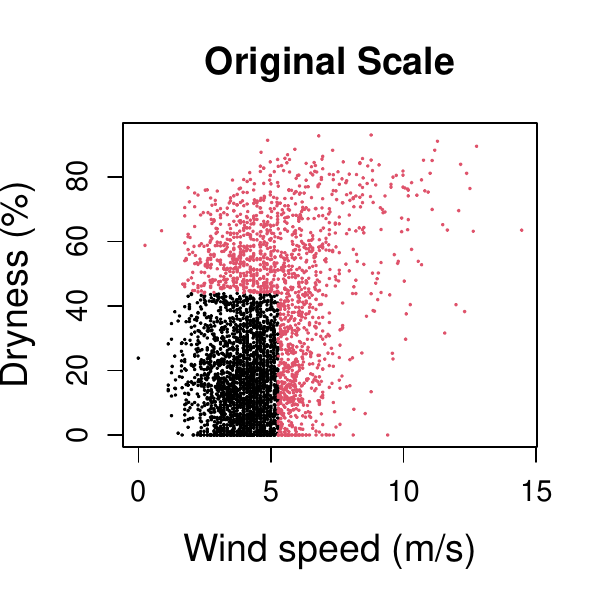}
    \includegraphics[width=0.49\textwidth]{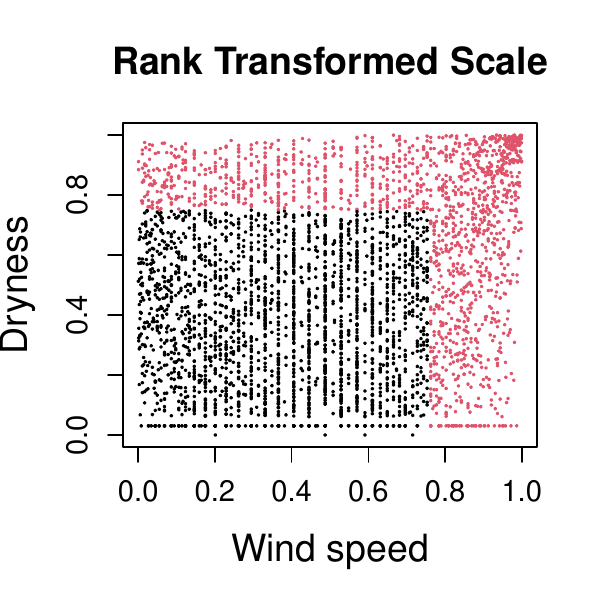}

\caption{Santa Ana wind speeds and dryness measured at the March Air Force Base station. Data above the $0.75$ marginal quantile threshold are in red.}
\label{fig:SantaAna_data}
\end{figure}

We apply our method to the Santa Ana winds and dryness data \citep{cooley2019}. The Santa Ana winds are a multivariate meteorological regime that has been implicated as a major driver of large wildfires in southern California \citep{Billmire2014-px}. Wildfires are related to several conditions like temperature, humidity, wind speed, and fuel supply \citep{LittellEtAl2018}. Historically, the autumn months of September, October, and November have had a higher number of wildfires compared to the winter months, and are associated with warm temperatures, low humidity, and high winds. \cite{cooley2019} surmised that the data exhibits AD and used the framework of regular variation to estimate probabilities associated with two different risk regions. The regular variation structure employed by them, however, cannot capture the nuance of AI. Our analysis could produce more accurate estimates of the probabilities of joint tail events that require extrapolation beyond the range of the data. In particular, if the data were asymptotically independent, an assumption of AD would overestimate the probability of such jointly extreme events under the approach used by \cite{cooley2019}.

We consider daily dryness (\%) and wind speed (m/s) data collected at the March Air Reserve Base station in Riverside County from the HadISD dataset \citep{HadISD}. The dryness is defined in this case as the $100-$RH, where RH is the relative humidity measured as a percentage. The bivariate time series represents a measure of the daily risk of fire. The station's variables have appeared to be associated with known Santa Ana events. The data consists of 3,902 days for the months of September--November from 1972--2015; we assume temporal stationarity and independence as in \cite{cooley2019}. Figure \ref{fig:SantaAna_data} plots the data both in its original scale as well as the rank transformed scale. The data shows tail dependence, noticeable in the rank transformed data with a cluster of values in the upper right corner. Our goal is to study the tail dependence between the two variables by estimating a gauge function for the data after a further transformation to unit exponential margins.

We analyze this data at a threshold of $\tau = 0.75$, providing us with 1,529 points that are above the threshold in at least one margin. The conditional distribution of the radii are assumed to be a truncated gamma distribution for this analysis. We run 2 MCMC chains of 11,000 iterations each, discarding the first 1,000 from each chain as burn-in. The priors and the remainder of the MCMC settings are identical to the simulation study.

\begin{figure}
\centering
     \includegraphics[width=0.45\textwidth]{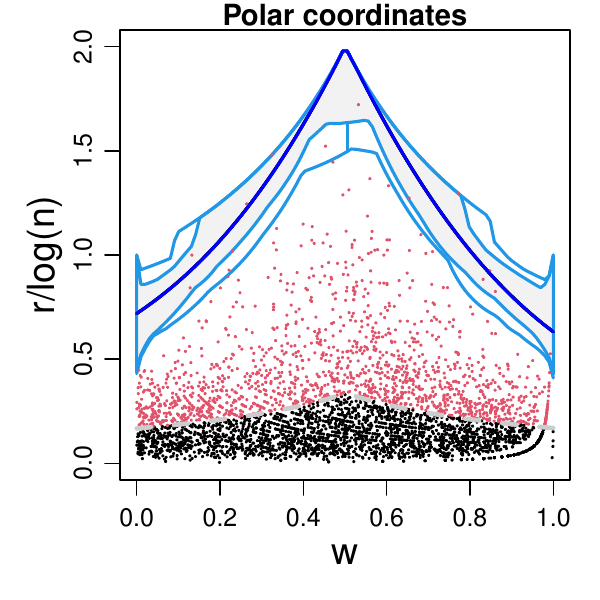}
    \includegraphics[width=0.45\textwidth]{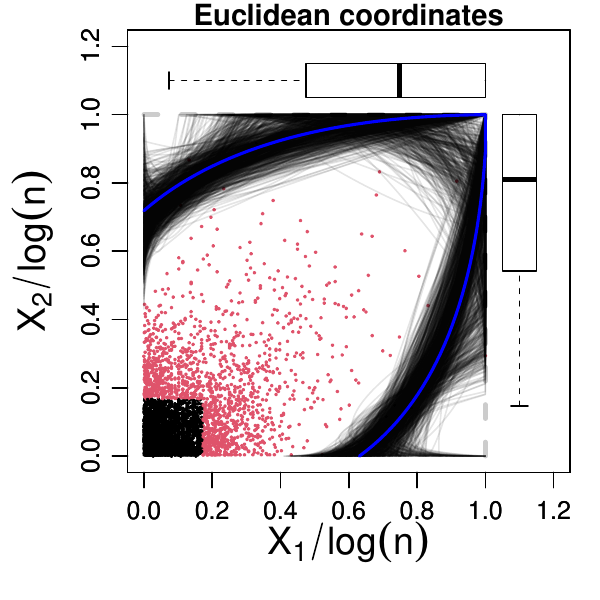}
\caption{Limit set boundaries based on \Bezier splines in pseudo-polar space (left) and Euclidean space (right) for assessing tail dependence between Santa Ana windspeed ($X_1$) and dryness ($X_2$). Median curves are plotted in dark blue.}
\label{fig:SantaAna_gauge}
\end{figure}
\begin{figure}
\centering
     \includegraphics[width=0.45\textwidth]{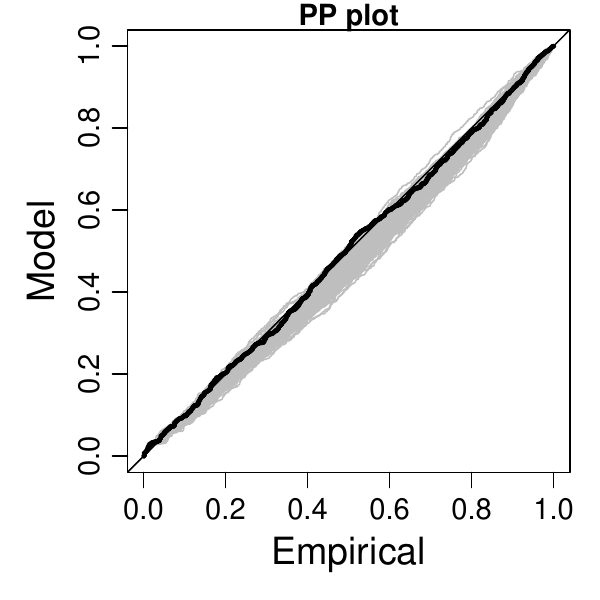}
    \includegraphics[width=0.45\textwidth]{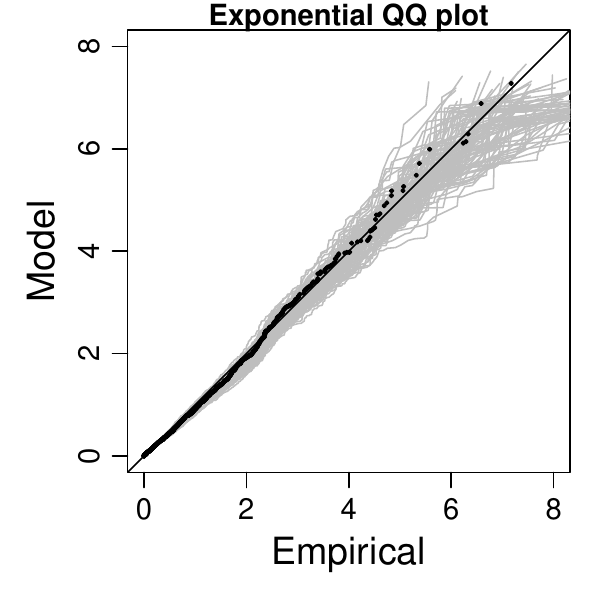}
\caption{PP plot (left) and exponential QQ plot (right) for the truncated gamma model fitted to the Santa Ana winds data. The black points correspond to the fit for the median \Bezier spline. Gray lines are based on 100 random draws from the posterior.}
\label{fig:SantaAna_modelfit}
\end{figure}
Figure \ref{fig:SantaAna_gauge} plots the estimated limit set boundaries, with the median curve plotted in blue. The plot on the left is in pseudo-polar coordinates and depicts a functional boxplot of the estimated limit set boundary, while the plot on the right is in Euclidean coordinates. The posterior median of $\eta$ is estimated to be 1, and $\mbox{P}(\eta=1 \given \bX) = 0.60$, suggesting AD between wind speed and dryness. Figure \ref{fig:SantaAna_modelfit} evaluates the goodness-of-fit for the truncated gamma model in terms of PP plots as well as QQ plots, based on post burn-in MCMC samples $\alpha^*$ and $\btheta^*_g$ from the posterior distribution of the parameters. 
The PP plot for $n_0$ observations with $R_i>r_0(\bw_i)$ is the set of points $[i/(n_0+1),u_{(n_0-i+1)}]$, where $u_{(1)}\geq u_{(2)}\geq \ldots \geq u_{(n_0)}$ is the ordered sample of $u_i = \hat{F}_t\bigl(r_i;\bw_i,r_0(\bw_i)\bigr)$, and $\hat{F}_t$ is the fitted CDF of the truncated gamma model with the likelihood function as in \eqref{eqn:data-likelihood}:
\[ 
 \hat{F}_t\bigl(r \vert \bw,r_0(\bw)\bigr) := \mbox{Pr}\bigl(R\leq r \vert W=\bw,R>r_0(\bw)\bigr) = 1-\frac{1 - F(r; \alpha^*, g_{\btheta^*_{g}}(\bw))}
            {1 - F(r_0(\bw); \alpha^*, g_{\btheta^*_g}(\bw))}.
\]
For the corresponding QQ plot, each point on the PP plot is transformed to unit exponential margins through an inverse CDF transformation. In both cases, the black points are obtained from gamma parameters implied by the median curve of the limit set boundary, while the gray lines correspond to $n_0=100$ random draws from the posterior. Both plots suggest that the model fits the data adequately.

To test sensitivity to the threshold level, we repeated the experiment at additional threshold levels of $\tau = 0.70, 0.80, \mbox{ and } 0.90$. In all 3 additional cases, $\mbox{P}(\eta=1 \given \bX) > 0.50$, with the smallest value (of 0.56) occurring for $\tau = 0.90$ and the largest value (of 0.64) for $\tau = 0.70$. Taken together, we conclude that dryness and windspeed for the Santa Ana dataset is asymptotically dependent with high probability, consistent with the results of \citet{cooley2019}.

\subsection{Analysis of Ozone concentration data}\label{s:Ozone}
For our second study, we consider air pollution measurement data across the US from the Community Multiscale Air Quality (CMAQ) model \citep{Binkowski2003,APPEL20079603,WYATAPPEL20086057} as well as EPA Air Quality System (AQS) \citep{AQS} data for the contiguous US. While CMAQ is a numerical model available across the entire country at a 12km resolution, the AQS dataset consists of observations monitored at $1,376$ stations across the US. Among them, only 519 stations had over 600 observations, which is what we use for this analysis. The full dataset has previously been used by \citet{Gong2021-sk} to develop a combined data product for 12 air pollutants. When fusing data products, it is important to calibrate the model data to ground truth. In our application, we will verify how strong the dependence is between the AQS and CMAQ datasets for ozone, one of the 12 pollutants made available by both datasets.

Our data consists of daily ozone readings for the months of July--September from 2010--2014, resulting in a bivariate time series of CMAQ and AQS data for up to 610 days at each station. The sample correlations between the AQS and CMAQ data for the 519 stations range from 0.29--0.86 with a median of 0.69, suggesting a high level of agreement in the bulk of the distribution. To assess tail dependence, we fit a gauge function with truncated gamma likelihood for data censored at $\tau=0.75$ threshold independently at every station. We run 2 MCMC chains for 11,000 iterations each for each station's data, discarding the first 1,000 as burn-in. 
\begin{figure}
\centering
     \includegraphics[width=0.45\textwidth]{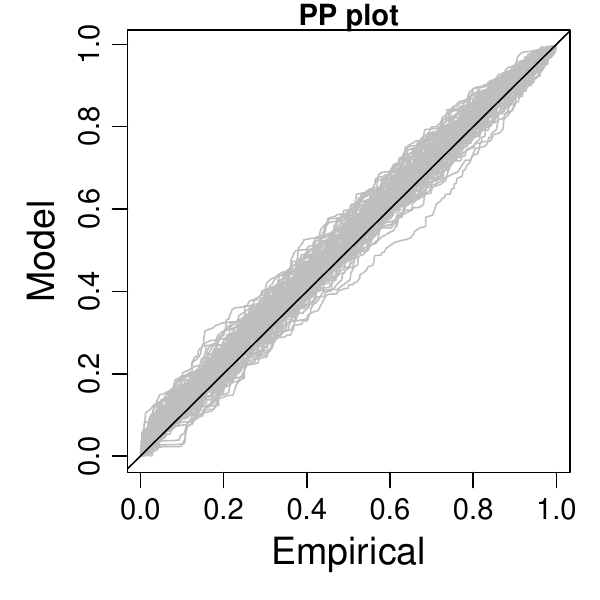}
    \includegraphics[width=0.45\textwidth]{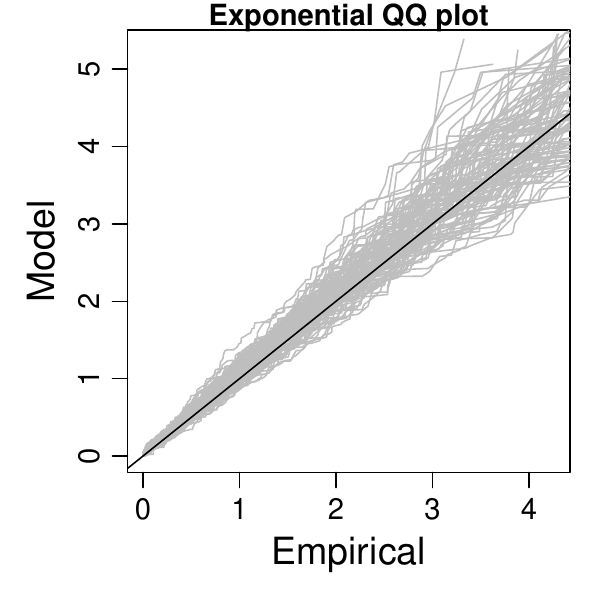}
\caption{PP plot (left) and exponential QQ plot (right) for the truncated gamma model fitted to the Ozone concentration data. The gray lines correspond to the fit for the median \Bezier splines at 100 random locations.}
\label{fig:CMAQ_modelfit}
\end{figure}

\begin{figure}
    \centering
    \includegraphics[width=\textwidth]{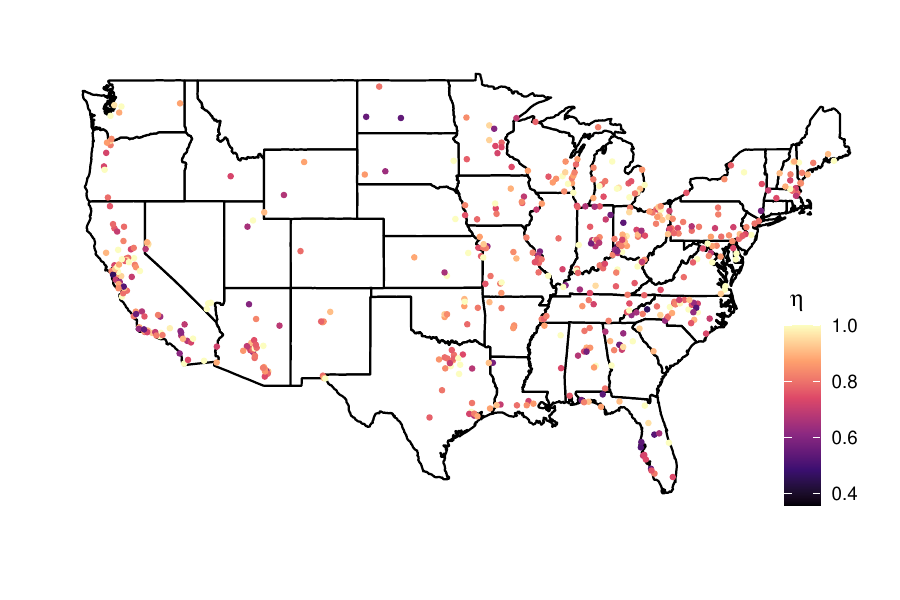}
    \caption{Posterior median of $\eta$ at 519 AQS monitoring stations in the US.}
    \label{f:eta_map}
\end{figure}

The posterior median of $\eta$ has an average value of 0.81 across the 519 locations, and is 1 (asymptotically dependent) for 79 of those stations. This suggests that the CMAQ data product can adequately represent the tail behavior of observational ambient ozone. Figure \ref{f:eta_map} plots the posterior probability of AD based on the truncated gamma model. While we are unable to discern any spatial pattern for high or low posterior values of $\eta$ from the map, we do note that several of the low values are in urban areas with high population densities. 

Finally, to study the sensitivity of our posterior to the threshold level $\tau$, we repeated our analysis at two additional values of $\tau = 0.70 \mbox{ and } 0.80$. Estimates from both these cases were quite similar to the baseline case of $\tau=0.75$, with correlations $>0.90$ for both the posterior median of $\eta$ and $\mbox{P}(\eta = 1)$. There were 76 and 80 locations respectively where the posterior median of $\eta$ was 1, and 62 of those locations were shared with the baseline case. Thus, our results are not very sensitive to the choice of threshold for small data sizes.
\section{Discussion}
Key aspects of tail dependence in multivariate distributions can be described through their corresponding gauge functions. In this study, we propose a semiparametric method for estimating gauge functions by modeling their unit level sets as \Bezier splines comprised of three quadratic \Bezier curves. The splines can represent the gauge function unit level sets, and hence limit set boundaries, of varying shapes, and are parsimoniously parameterized by a small number of control points. The quadratic specification makes it straightforward to obtain analytical solutions for the shape of the limit set, and constraints on the control points ensure that the resultant shapes are valid limit set boundaries.  Bayesian estimation of the \Bezier splines requires only standard MCMC techniques and allows important cases on the edge of the parameter space to be represented by employing mixture priors with point masses.  We demonstrate the efficacy of our model using numerical studies as well as two real data applications involving fire weather in California, and ambient air pollution from ozone across the US.

We have only considered bivariate random vectors here, but the modeling strategy can scale to three dimensions by using \Bezier surfaces, with the control points of constituent \Bezier curves set in $\mathbb{R}^3$ instead of $\mathbb{R}^2$. It will, however, require more complicated constraints to ensure that the star shaped property holds, and dimensions greater than three appear to be infeasible. 
In addition, it appears possible to extend our modeling framework to include negative dependence by transforming to Laplace margins rather than exponential margins.  This has been previously suggested \citep{simpson-2022a,wadsworth-2022a} and recently implemented in the context of radially stable Pareto distributions \citep{papastathopoulos-2023}, and within a semiparametric angular-radial model \citep{mackay2023modelling}.  Implementing our \Bezier model in Laplace margins (in two dimensions) would require somewhere between two to four times the number of control points as we have now, specifying appropriate constraints on their support, and more sophisticated sampling algorithms to ensure convergence.
Finally, while three \Bezier curves are sufficient to ensure that the boundary of our estimated limit set in the two dimensional case exactly touches the corners of the unit box, the curves themselves don't necessarily need to be quadratic. It is possible to use a nonparametric Bayesian framework to construct curves that have an arbitrary number of intermediate control points (i.e., control points excluding the start and end points). Though this would be computationally more expensive, the resulting \Bezier spline is likely to have broader support over the space of all limit set boundaries.
\section*{Funding}
This work was supported by grants from the Southeast National Synthesis Wildfire and the United States Geological Survey’s National Climate Adaptation Science Center (G21AC10045), and the National Science Foundation (DMS-2001433, DMS-2152887). We acknowledge the computing resources provided by North Carolina State University High Performance Computing Services Core Facility (RRID:SCR\_022168).
\bibliographystyle{asa}
\bibliography{sources}

\end{document}


\DeclareGraphicsExtensions{.png}

\begin{center}
\textbf{\LARGE Supplementary Material for `Semiparametric Estimation of the Shape of the Limiting Bivariate Point Cloud'} \\
\onehalfspacing
\large Reetam Majumder, Benjamin A. Shaby, Brian J. Reich, and Daniel S. Cooley \\
\onehalfspacing
\end{center}

\singlespacing
\begin{appendix}

\section{Comparison of Threshold Selection Approaches}
\subsection{Connection to the main text}
This simulation study supports Section 3.1 of the main text, and investigates the sensitivity of the form of threshold used to select data from the joint tail on the estimation of $\eta$. 
\subsection{Different forms of the threshold}

Previous works \citep{wadsworth-2022a,simpson-2022a} have used elaborate schemes to specify a good threshold as part of their pipelines for estimating the gauge function.  These schemes all in essence come down to some version of estimating a conditional high quantile of radii, as a function of angle.  However, both the truncated and censored gamma likelihoods that we use here contain terms that account for the magnitude of the threshold at any given angle, so it is not obvious to us why choosing a complicated threshold is necessary for estimating the gauge function.  To explore this issue, we compare two schemes for choosing the threshold, which range from the simplest available choice to the unattainably ideal.

The first scheme is the \emph{marginal threshold}, which sets a threshold in each margin, independently, at a fixed marginal empirical quantile $\tau$. This is the approach that has been used in this study. The second scheme is the \emph{oracle threshold}.  This threshold is based on the observation that the conditional quantile is inversely proportional to the gauge function, evaluated at any angle, at high quantiles \citep{wadsworth-2022a}.  That is,
\[
  q_{\tau}(w) \approx \frac{C}{g(w)},
\]
for some constant $C$, for large $\tau$.  In the context of a simulation, we know the gauge function $g(w)$ of the data-generating model, so for any simulated dataset we can choose $C_0$ such that the proportion of radii that exceed $\frac{C_0}{g(w)}$ is $1-\tau$.  We then set the threshold $r_0(w) = \frac{C_0}{g(w)}$.  This threshold is therefore an asymptotic approximation to the true conditional quantile $q_\tau(w)$.  We call it the oracle threshold because it requires knowledge of the true gauge function $g(w)$, which is of course unavailable outside of simulations.  One can think of it as a fairly good but unknowable conditional quantile estimate which captures the general shape of high conditional quantiles.

To put both thresholds on even footing for the purposes of gauge function estimation, we force them to include the same number of points available for estimation.  Let the 2 variables under consideration be $X_1$ and $X_2$; we denote their marginal quantiles as $q_1$ and $q_2$, where the quantiles are selected at the $\tau=0.75$ level. Further, let $n$ denote the number of observations in our study. The \textit{marginal threshold} that we use in our study selects $(X_1,X_2)$ points for which either $X_1>q_1$, or $X_2>q_2$, or both. The number of points that are selected above the threshold using this approach is of course higher than $n(1-\tau)$; this is denoted $n_0$ in Eqn. (2) of the main text. We then adjust the nominal quantile level $\tau$ when calculating the radial and oracle thresholds, such that each results in $n_\text{marg}$ points available for model fitting.

\subsection{Results}
We compare the the two thresholds for the four copulas and across five dependence levels that are considered in the simulation study (Section 4) in the main text. For each of the twenty configurations, we generate 100 datasets of $n=5,000$ data points. The analysis based on the marginal threshold is identical to the simulation study presented in  the main text, at a quantile level of $\tau=0.75$. We also extract the number of points used for fitting the truncated Gamma likelihood for the 100 datasets. This number is different for each of the 100 datasets, and we denote them $n_{0}^{(1)},\ldots, n_{0}^{(100)}$. To evaluate model fit based on the oracle thresholds, we select $n_{0}^{(1)},\ldots, n_{0}^{(100)}$ points respectively from the 100 datasets. This ensures that the truncated Gamma likelihood is fitted to the same number of points in each case. The MCMC simulations are carried out using the same settings as the simulation study in the main text.

\begin{table}
\centering
\small
\caption{RMSE ratios for estimates of $\eta$ and RMISE ratios for estimates of $\lambda(\omega)$ and $\tau_1(\delta)$ based on the marginal $(\hat{\eta}_m,\hat{\lambda}_m,\mbox{ and }\hat{\tau}_m)$ and oracle $(\hat{\eta}_o,\hat{\lambda}_o,\mbox{ and }\hat{\tau}_o)$ thresholds over simulated datasets for four copulas and five dependence levels.}
\label{t:rmise_thresh}
\begin{tabular}{ccccccc}\toprule
\textbf{Measure} & \textbf{Copula} & \multicolumn{5}{c}{\textbf{Dependence parameter value}} \\
 &  & 0.3 & 0.4 & 0.5 & 0.6 & 0.7 \\\midrule
\multirow{4}{*}{$\frac{RMSE(\hat{\eta})}{RMSE(\Tilde{\eta})}$} & Gaussian & 1.09  &  1.03 & 0.92 & 1.00 & 0.93 \\
 & Logistic & 0.90 & 1.21 & 0.91 & 1.12 & 0.96 \\
 & Inv-Logistic & 0.90 & 1.11 & 1.03 & 0.98 & 0.87 \\
 & Asy-Logistic & 1.27 & 0.93 & 1.05 & 0.96 & 1.44 \\\midrule
\multirow{4}{*}{$\frac{RMISE(\hat{\lambda})}{RMISE(\Tilde{\lambda})}$} & Gaussian & 1.09 & 1.06 & 0.96 & 0.99 & 0.98 \\
 & Logistic & 0.89 & 1.17 & 0.88 & 1.17 & 0.76 \\
 & Inv-Logistic & 0.99 & 1.06 & 1.02 & 1.00 & 0.94 \\
 & Asy-Logistic & 1.24 & 1.01 & 0.33 & 0.65 & 1.61 \\\midrule
\multirow{4}{*}{$\frac{RMISE(\hat{\tau})}{RMISE(\Tilde{\tau})}$} & Gaussian & 1.04 & 1.06 & 1.03 & 1.02 & 1.02 \\
 & Logistic & 1.11 & 1.05 & 1.05 & 1.04 & 1.01\\
 & Inv-Logistic & 1.09 & 1.25 & 0.95 & 1.00 & 1.05 \\
 & Asy-Logistic & 0.73 & 0.93 & 0.76 & 0.70 & 0.54 \\\bottomrule
\end{tabular}
\normalsize
\end{table}

\begin{table}
\small
\centering
\caption{Number of datasets (out of 100), where the posterior median of $\eta$ based on the marginal threshold is 1 for each scenario. Values in parenthesis are for estimates based on the oracle threshold.}
\label{t:eta_estimates_thresh}
\begin{tabular}{cccccc}\toprule
 & \multicolumn{5}{c}{\textbf{Dependence parameter value}} \\
 & 0.3 & 0.4 & 0.5 & 0.6 & 0.7 \\\midrule
Gaussian & 00 (00) & 00 (00) & 00 (00) & 00 (00) & 06 (03) \\
Logistic & 98 (97) & 93 (92) & 96 (96) & 91 (88) & 82 (78) \\
Inv-Logistic & 02 (00) & 00 (00) & 00 (00) & 00 (00) & 00 (00) \\
Asy-Logistic & 96 (93) & 80 (85) & 82 (80) & 83 (81) & 84 (74)\\\bottomrule
\end{tabular}
\normalsize
\end{table}

Table \ref{t:rmise_thresh} contains RMSE ratios for estimates of $\eta$ and RMISE ratios for estimates of $\lambda(\omega)$ and $\tau_1(\delta)$ based on the marginal and oracle thresholds. For estimates of $\eta$ and $\lambda(\omega)$, the marginal threshold has lower error in half the cases, while for $\tau_1(\delta)$, the marginal threshold has lower errors in $70\%$ of cases. This suggests that using the marginal threshold does not noticeably affect the estimation of the metrics under consideration. Table \ref{t:eta_estimates_thresh} summarizes the number of datasets (out of 100) for each configuration where the posterior median of $\eta$ is 1. We find the marginal threshold to provide slightly better results for the AD copulas, and the oracle threshold to provide slightly better results for AI copulas.

\begin{figure}
    \centering
    \includegraphics[width=0.4\textwidth]{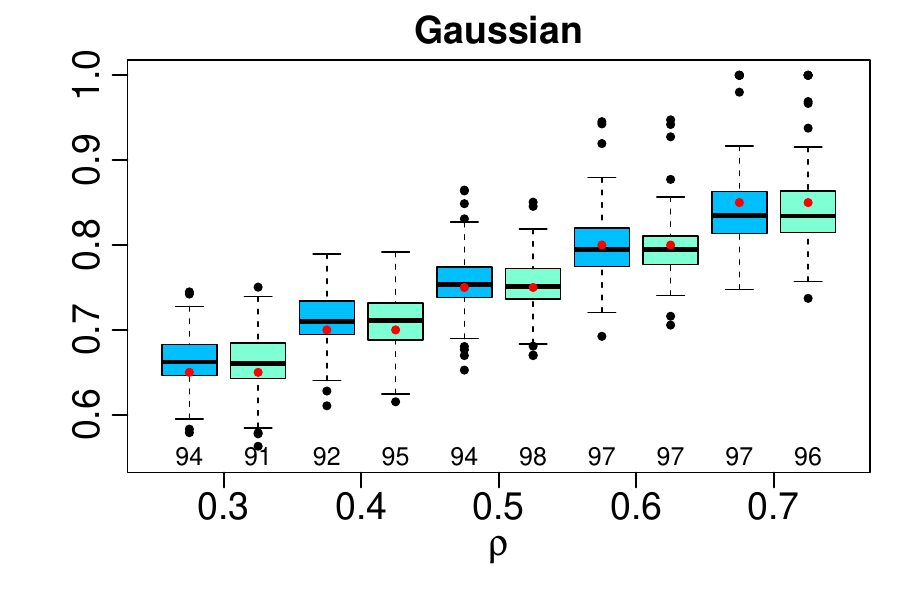}
    \includegraphics[width=0.4\textwidth]{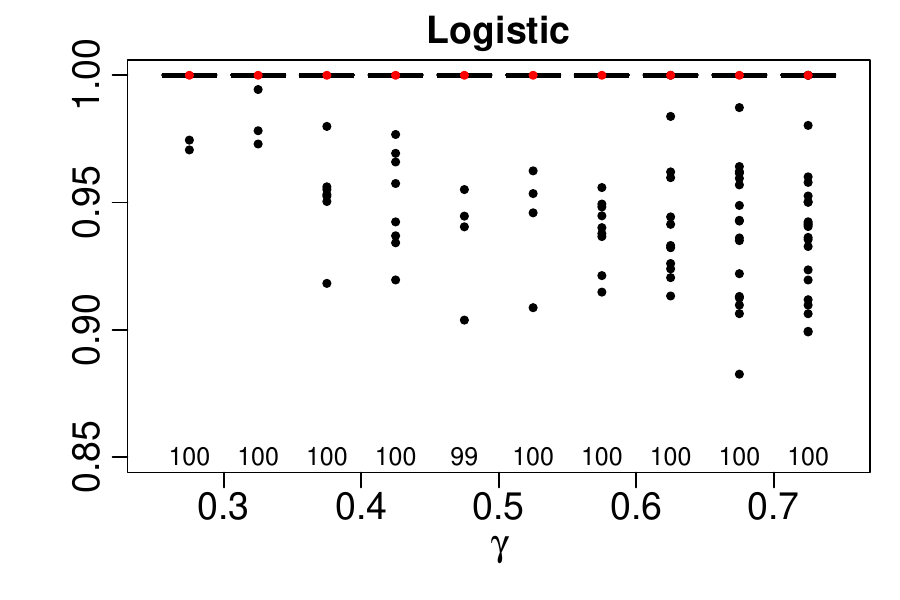}
    \includegraphics[width=0.4\textwidth]{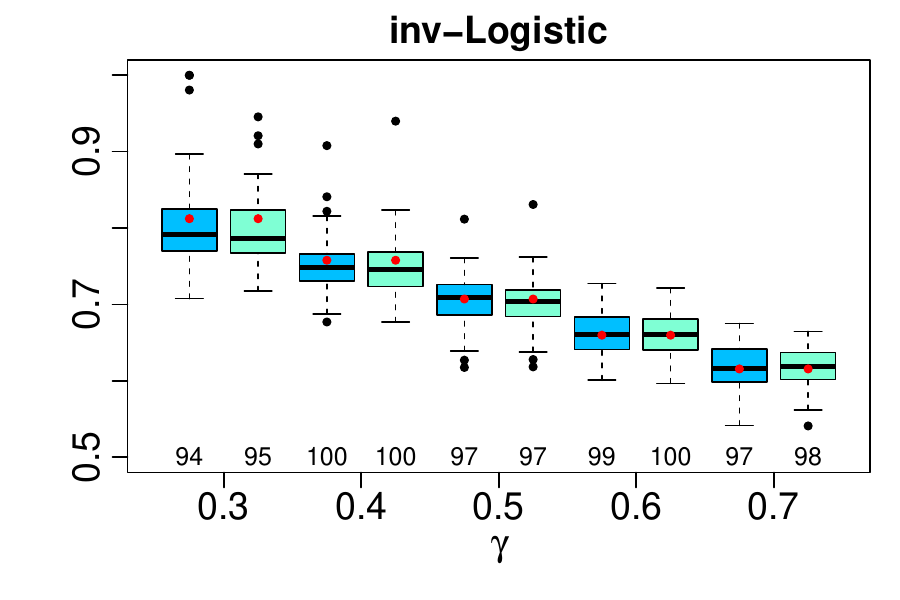}
    \includegraphics[width=0.4\textwidth]{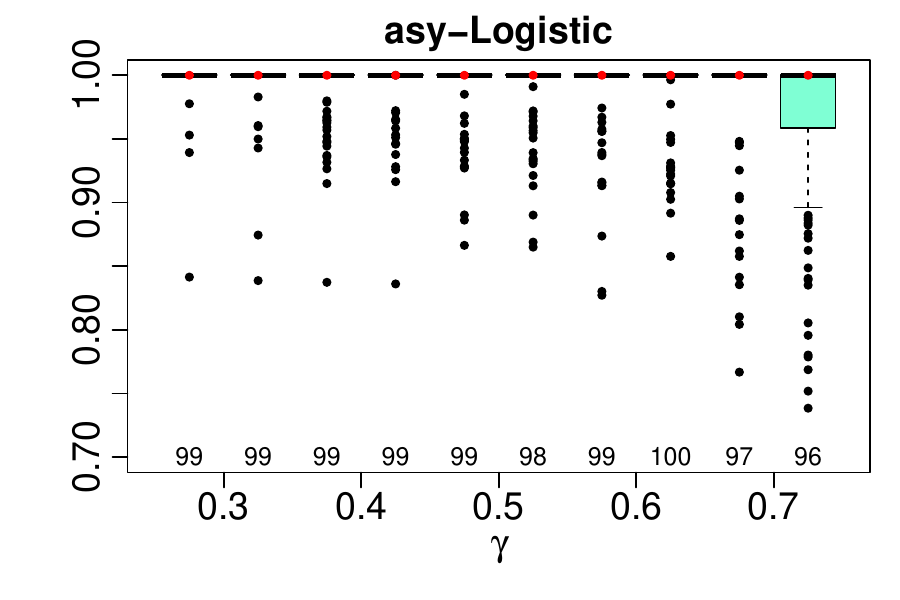}
    \caption{\textbf{Choice of threshold for the \Bezier spline estimator:} Sampling distribution of the posterior medians of $\eta$ based on the marginal (blue) and oracle (green) thresholds for the four dependence copulas. The red dots indicate the true values, and coverage of equi-tailed 95\% intervals are noted below each boxplot.}
    \label{f:MCMC_eta_thresh}
\end{figure}

\begin{figure}
    \begin{subfigure}{\textwidth}
    \centering
    \includegraphics[width=0.4\textwidth]{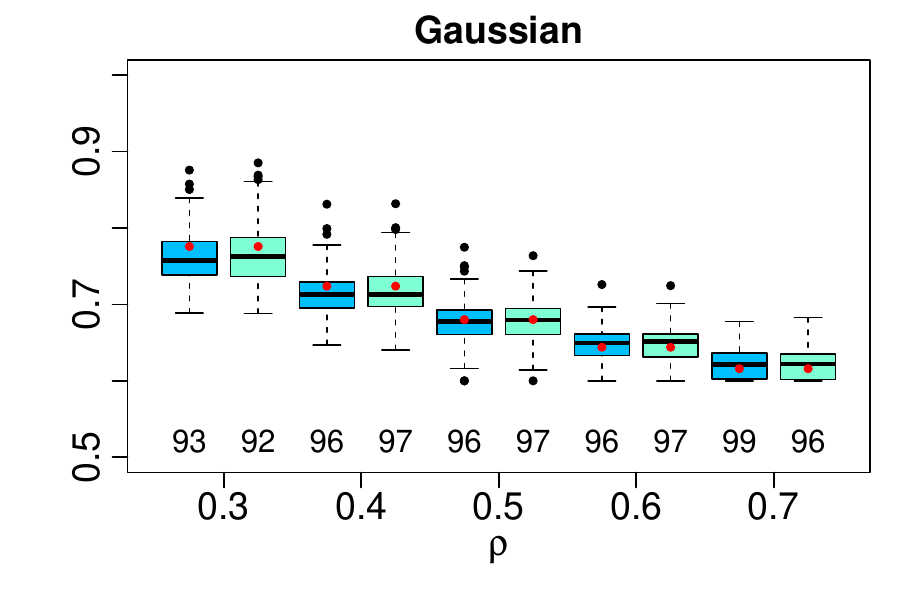}
    \includegraphics[width=0.4\textwidth]{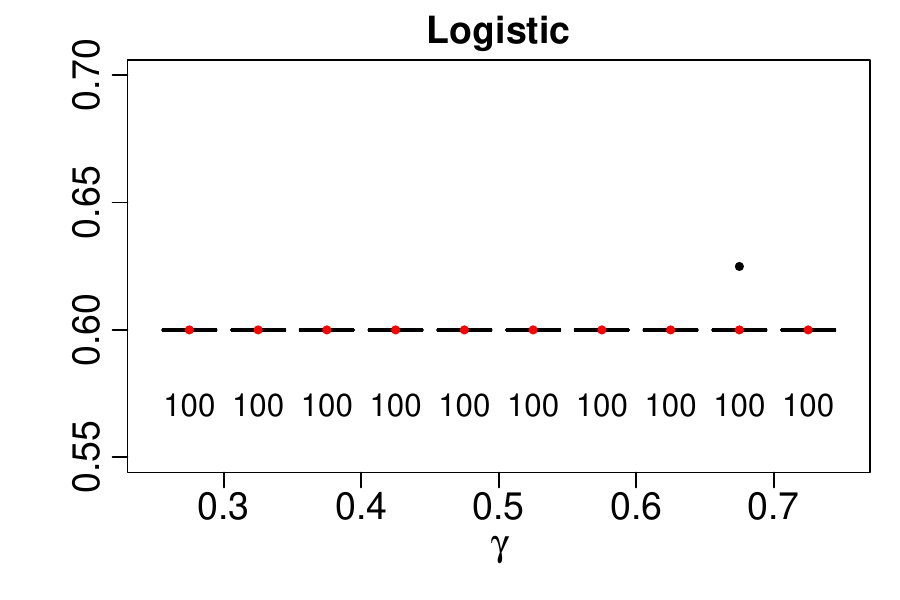}
    \includegraphics[width=0.4\textwidth]{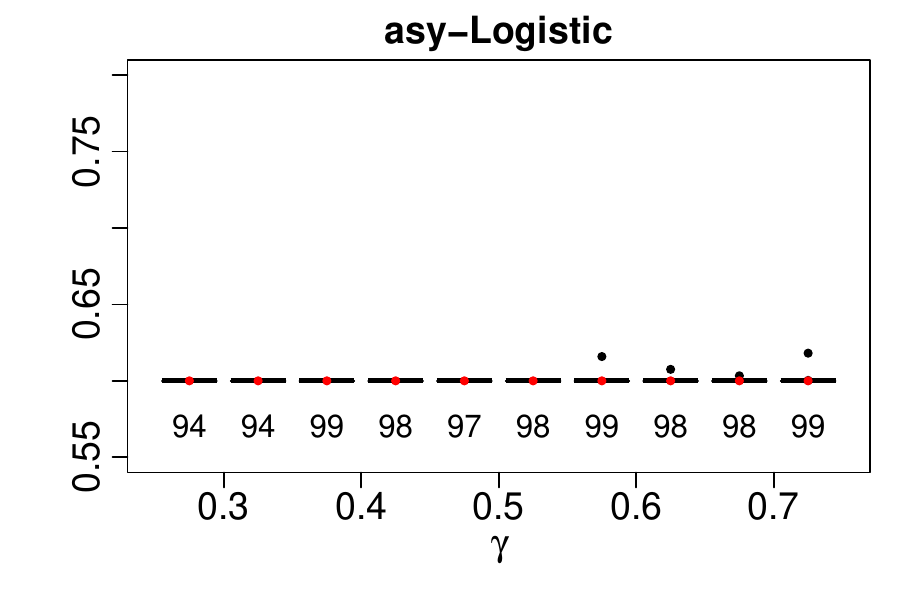}
    \includegraphics[width=0.4\textwidth]{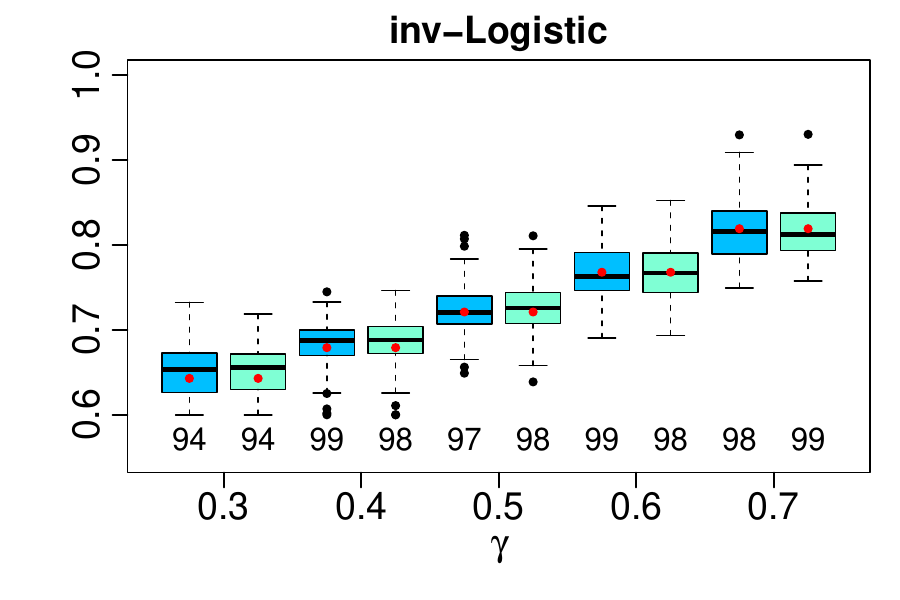}
    \caption{Sampling distribution of the posterior medians of $\lambda(0.40)$.}
    \label{f:MCMC_lambda_thresh}
    \end{subfigure}
\begin{subfigure}{\textwidth}
\centering
    \includegraphics[width=0.4\textwidth]{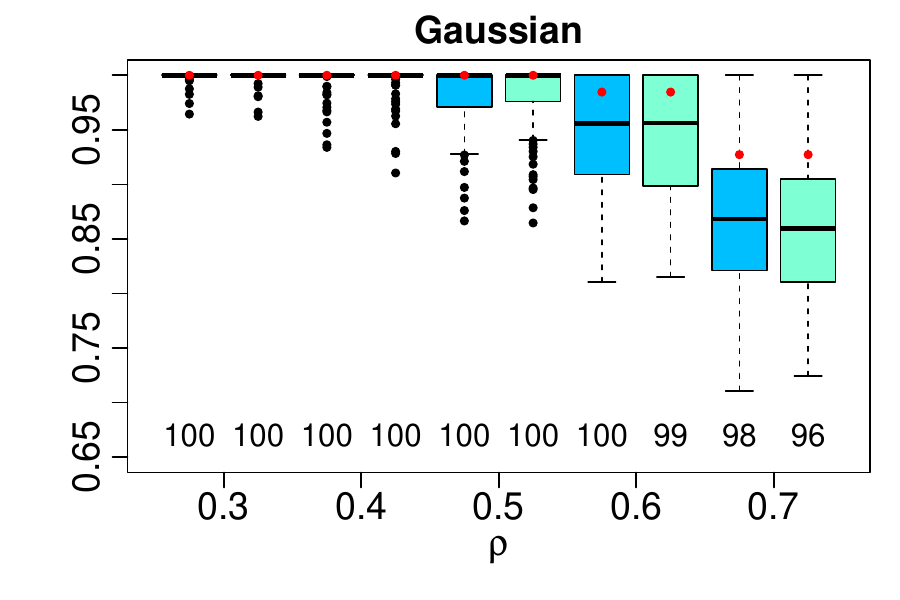}
    \includegraphics[width=0.4\textwidth]{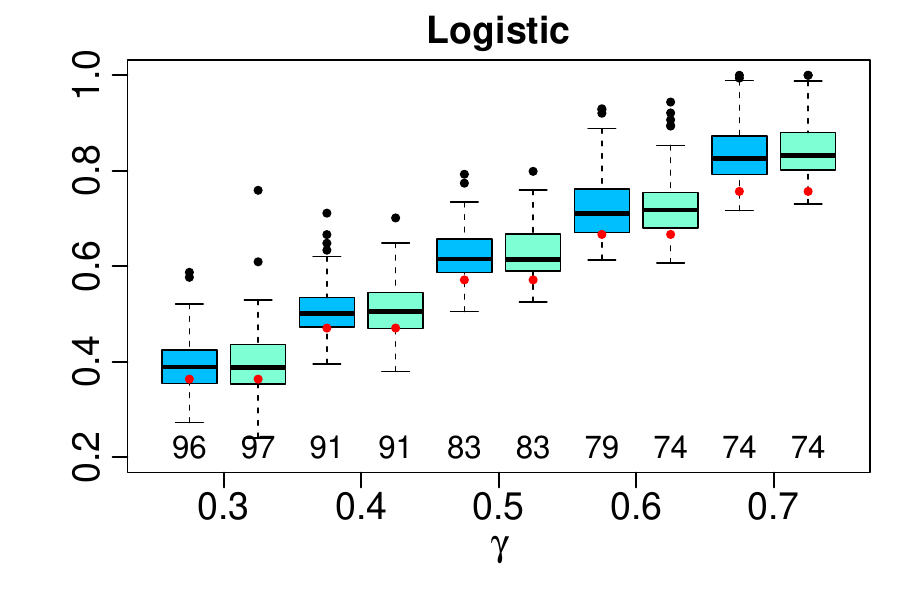}
    \includegraphics[width=0.4\textwidth]{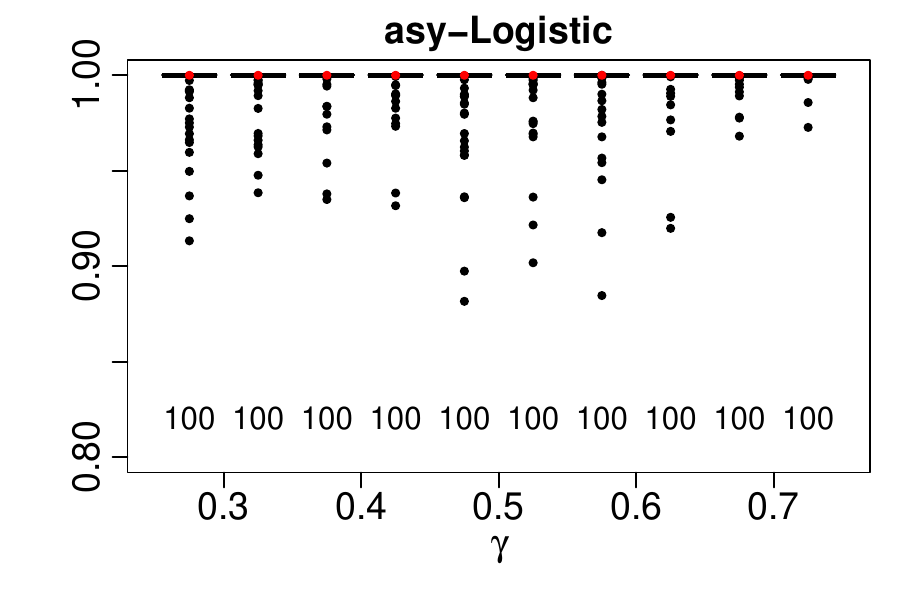}
    \includegraphics[width=0.4\textwidth]{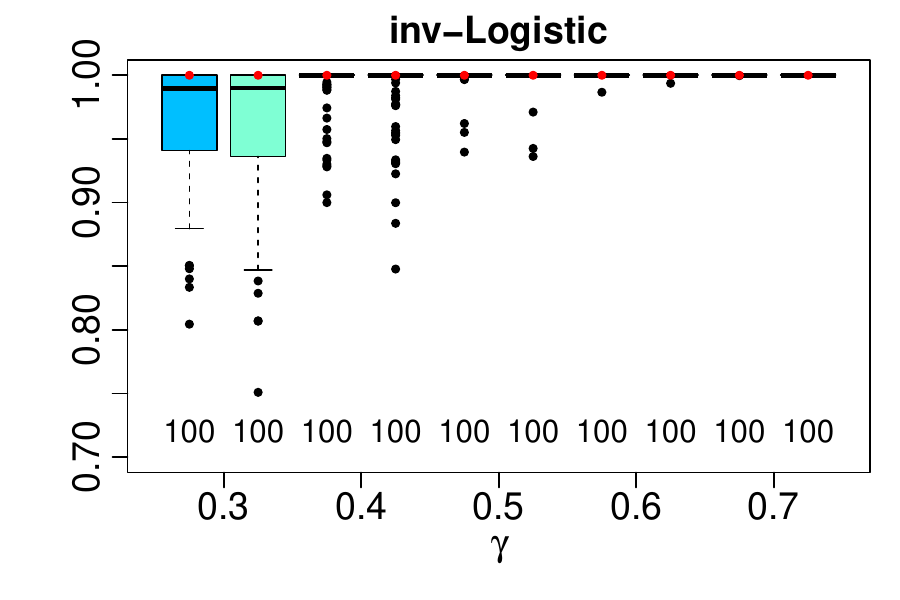}
    \caption{Sampling distribution of the posterior medians of $\tau_1(0.25)$.}
    \label{f:MCMC_tau_thresh}
    \end{subfigure}
     \caption{Sampling distribution of the posterior medians of $\lambda(0.40)$ and $\tau_1(0.25)$ based on the \Bezier spline (blue) and Simpson-Tawn (green) estimators for four dependence copulas, with dependence parameters set to 0.5. The red lines indicate the true values, and coverage of equi-tailed 95\% intervals are noted below each boxplot.}
    \label{f:MCMC_lambda_and_tau_thresh}
\end{figure}

Figure \ref{f:MCMC_eta_thresh} shows boxplots of the posterior median of $\eta$ for the different cases. Analytical values of $\eta$ are shown as red points in each plot, and the coverage of equi-tailed 95\% intervals are  noted below each boxplot. Results when using marginal thresholding are very similar to results when using oracle thresholding in all cases. Similar behavior is seen in Figure \ref{f:MCMC_lambda_and_tau_thresh}, which plots posterior medians of $\lambda(0.40)$ and $\tau_1(0.25)$ for the four copulas when the dependence parameter is 0.5. This indicates that the very simple marginal threshold performs comparably to the (unattainable) oracle threshold.  We thus proceed by eschewing complicated threshold calculation requiring quantile regression, and simply use the marginal threshold.
\clearpage
\section{Comparison of Estimators for Small Datasets}
\subsection{Connection to the main text}
This Appendix supports Section 4 of the main text, and compares estimates of $\eta, \lambda(\omega)$, and $\tau_1(\delta)$ based on the \Bezier spline and Simpson-Tawn estimators, for datasets of size 600.
\subsection{Results}
\begin{table}
\centering
\small
\caption{RMSE ratios for estimates of $\eta$ and RMISE ratios for estimates of $\lambda(\omega)$ and $\tau_1(\delta)$ based on the \Bezier spline $(\hat{\eta},\hat{\lambda},\mbox{ and }\hat{\tau})$ and Simpson-Tawn $(\tilde{\eta},\tilde{\lambda},\mbox{ and }\tilde{\tau})$ estimators over simulated datasets for four copulas and five dependence levels, based on a data size of 600.}
\label{t:rmise_600}
\begin{tabular}{ccccccc}\toprule
\textbf{Measure} & \textbf{Copula} & \multicolumn{5}{c}{\textbf{Dependence parameter value}} \\
 &  & 0.3 & 0.4 & 0.5 & 0.6 & 0.7 \\\midrule
\multirow{4}{*}{$\frac{RMSE(\hat{\eta}_o)}{RMSE(\hat{\eta}_m)}$} & Gaussian & 0.85 & 0.89 & 0.95 & 0.94 & 1.03 \\
 & Logistic & Inf & 1.70 & 0.81 & 0.83 & 1.21 \\
 & Inv-Logistic & 1.04 & 1.12 & 1.14 & 0.89 & 0.96 \\
 & Asy-Logistic & 3.18 & 2.67 & 2.25 & 1.53 & 1.52 \\\midrule
\multirow{4}{*}{$\frac{RMISE(\hat{\lambda}_o)}{RMISE(\hat{\lambda}_m)}$} & Gaussian & 1.06 & 1.18 & 1.27 & 1.28 & 1.21 \\
 & Logistic & Inf & 2.55 & 1.12 & 1.60 & 2.31 \\
 & Inv-Logistic & 1.19 & 1.17 & 1.15 & 1.12 & 1.06 \\
 & Asy-Logistic & 6.45 & 10.6 & 6.74 & 2.79 & 2.59 \\\midrule
\multirow{4}{*}{$\frac{RMISE(\hat{\tau}_o)}{RMISE(\hat{\tau}_m)}$} & Gaussian & 0.36 & 0.54 & 0.69 & 0.85 & 1.04 \\
 & Logistic & 0.76 & 0.85 & 0.90 & 1.11 & 1.26 \\
 & Inv-Logistic & 0.26 & 0.15 & 0.15 & 0.11 & 0.19 \\
 & Asy-Logistic & 0.66 & 0.68 & 0.68 & 0.73 & 0.33 \\\bottomrule
\end{tabular}
\normalsize
\end{table}

\begin{table}
\small
\centering
\caption{Number of datasets (out of 100), each of size 600, where the posterior median of $\eta$ is 1 for each scenario. Values in parenthesis correspond to the Simpson-Tawn estimator.}
\label{t:eta_estimates_600}
\begin{tabular}{cccccc}\toprule
 & \multicolumn{5}{c}{\textbf{Dependence parameter value}} \\
 & 0.3 & 0.4 & 0.5 & 0.6 & 0.7 \\\midrule
Gaussian & 01 (00) & 01 (00) & 02 (00) & 03 (00) & 06 (00) \\
Logistic & 100 (58) & 92 (62) & 77 (59) & 66 (47) & 51 (21) \\
Inv-Logistic & 07 (01) & 01 (00) & 00 (00) & 02 (00) & 01 (00) \\
Asy-Logistic & 96 (28) & 88 (32) & 76 (18) & 57 (10) & 42 (03)\\\bottomrule
\end{tabular}
\normalsize
\end{table}
Table \ref{t:rmise_600} summarizes the RMSE ratio for estimates of $\eta$ and RMISE ratios for estimates of $\lambda(\omega)$ and $\tau_1(\delta)$ based on the \Bezier spline and Simpson-Tawn estimators. We see very similar behavior as in the simulation study presented in the main text, with the \Bezier spline estimator outperforming the Simpson-Tawn estimator when estimating $\eta$ and $\lambda(\omega)$, with the opposite being true for estimates of $\tau_1(\delta)$. This is also reflected in Table \ref{t:eta_estimates_600}, which provides the number of datasets (out of 100) in each scenario where the posterior medians of $\eta$ is estimated to be 1. The values in the parentheses are corresponding point estimates of $\eta$ based on the Simpson-Tawn estimator. The \Bezier spline estimator has slightly higher error for the AI cases; however, the Simpson-Tawn estimator has significantly higher error when trying to identify AD for small datasets than for the larger datasets used in the simulation studies that are presented in the main text.

\begin{figure}
    \centering
    \includegraphics[width=0.4\textwidth]{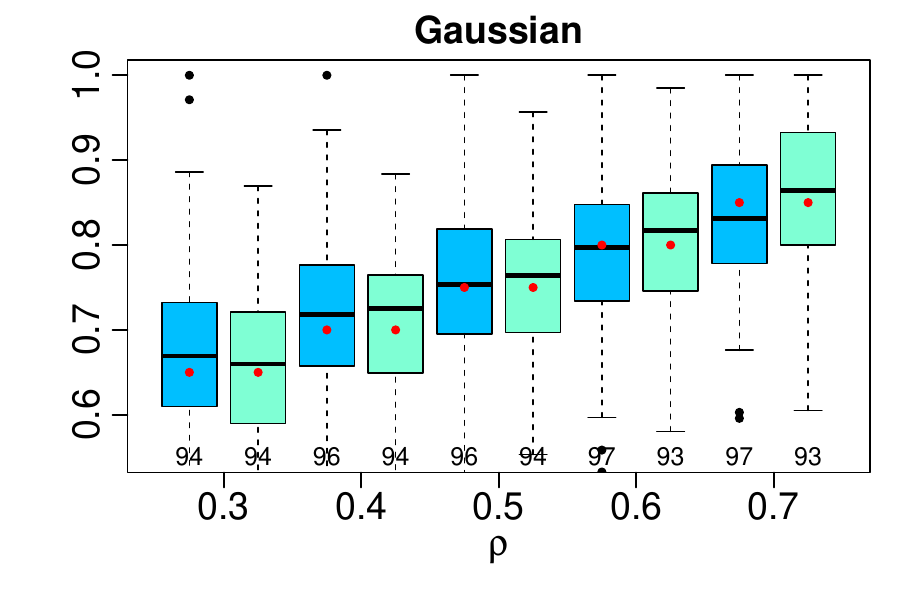}
    \includegraphics[width=0.4\textwidth]{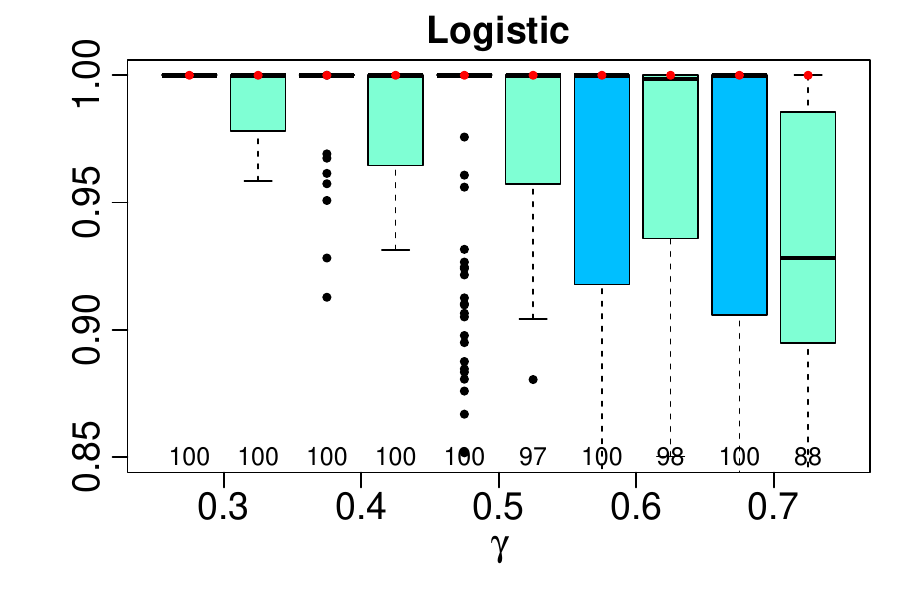}
    \includegraphics[width=0.4\textwidth]{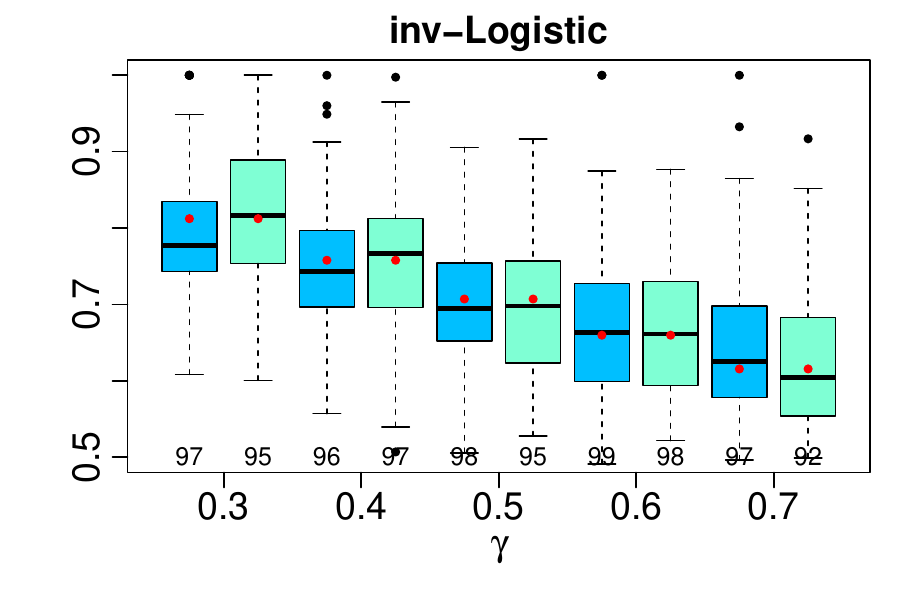}
    \includegraphics[width=0.4\textwidth]{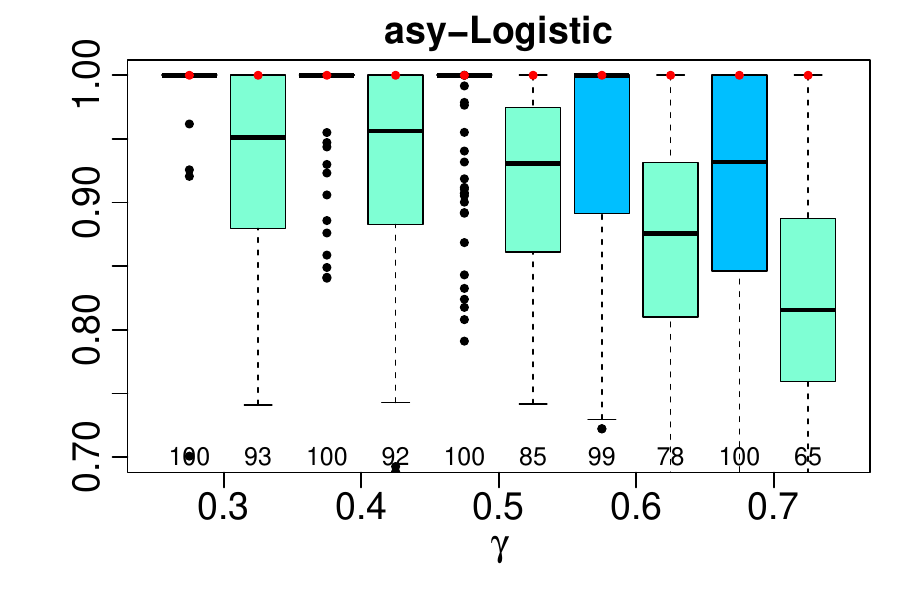}
    \caption{\textbf{Estimation of $\eta$ for a data size of 600:} Sampling distribution of the posterior medians of $\eta$ based on the \Bezier spline estimate (blue) for the four dependence copulas, alongside estimates using the Simpson-Tawn estimator (green). The red dots indicate the true values, and coverage of equi-tailed 95\% intervals are noted below each boxplot.}
    \label{f:MCMC_eta_600}
\end{figure}

\begin{figure}
    \begin{subfigure}{\textwidth}
    \centering
    \includegraphics[width=0.4\textwidth]{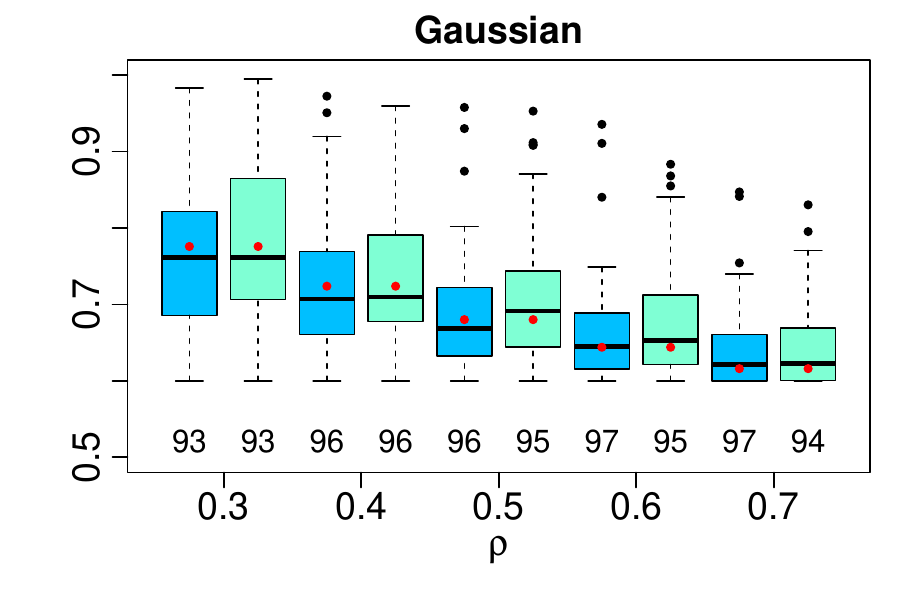}
    \includegraphics[width=0.4\textwidth]{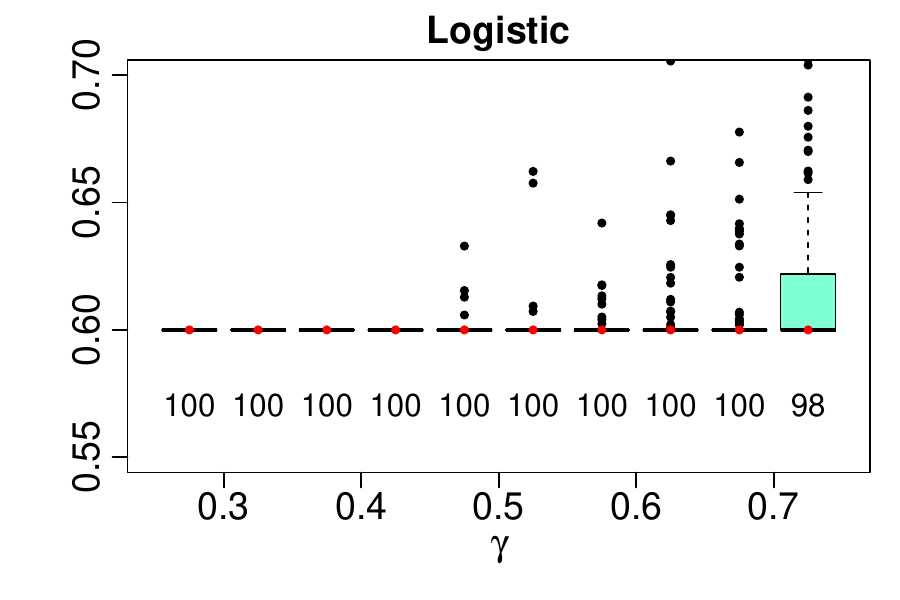}
    \includegraphics[width=0.4\textwidth]{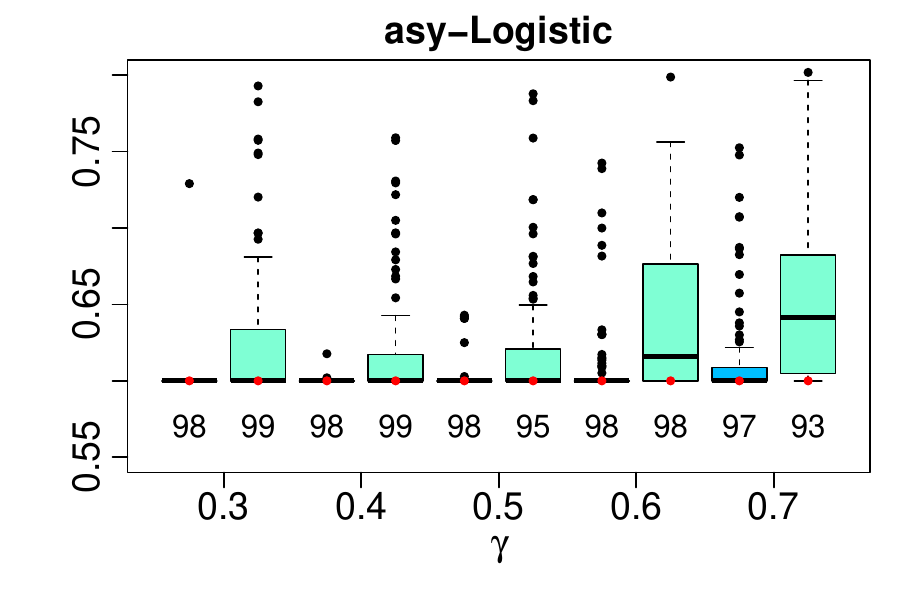}
    \includegraphics[width=0.4\textwidth]{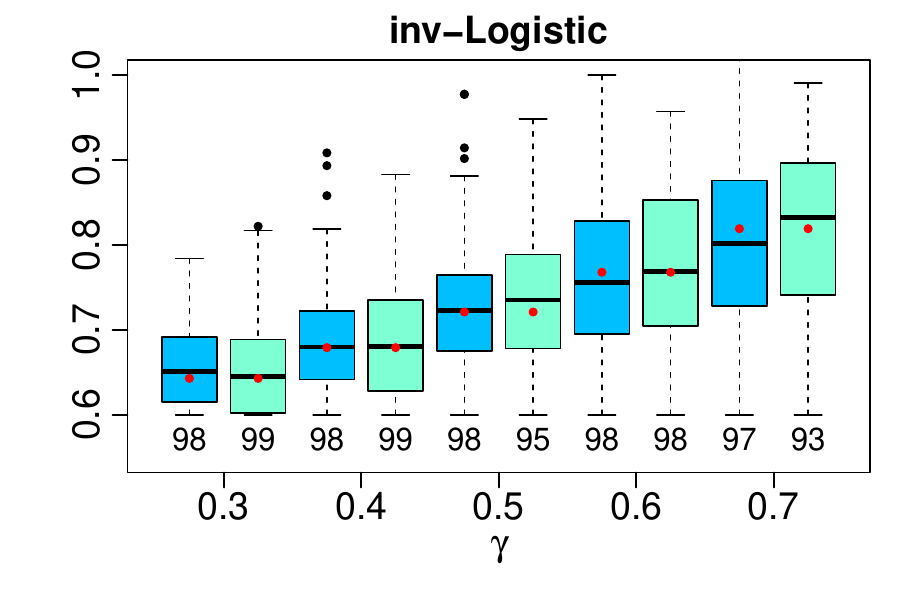}
    \caption{Sampling distribution of the posterior medians of $\lambda(0.40)$.}
    \label{f:MCMC_lambda_600}
    \end{subfigure}
\begin{subfigure}{\textwidth}
    \centering
    \includegraphics[width=0.4\textwidth]{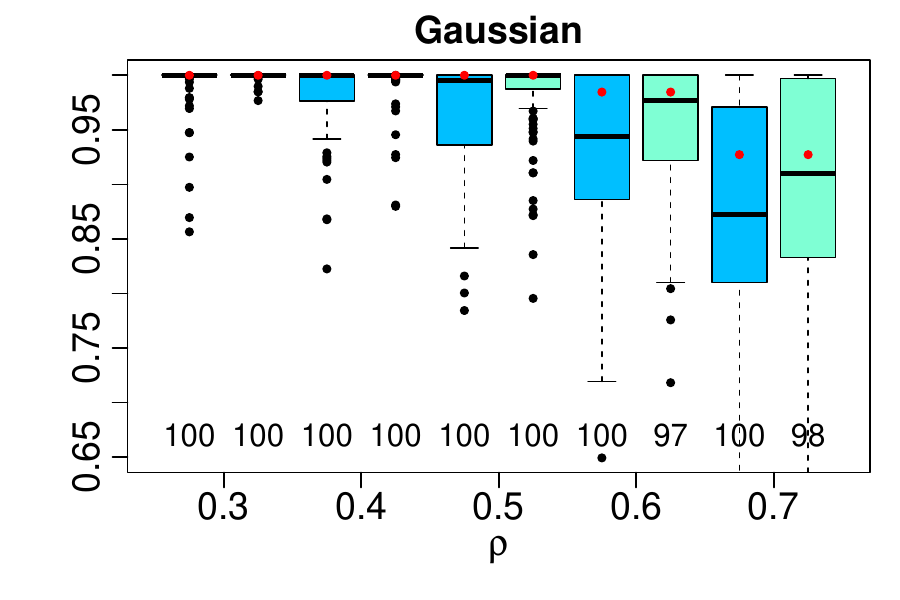}
    \includegraphics[width=0.4\textwidth]{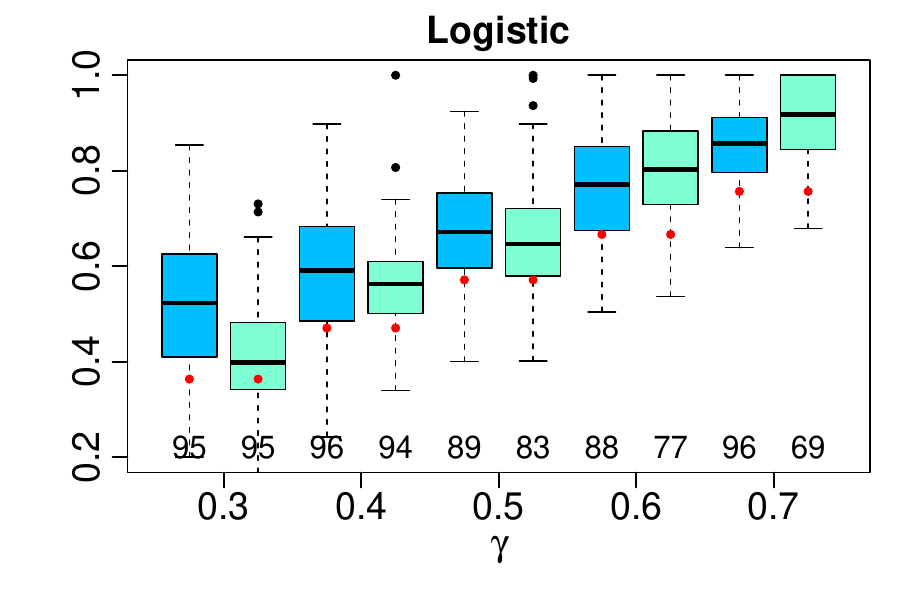}
    \includegraphics[width=0.4\textwidth]{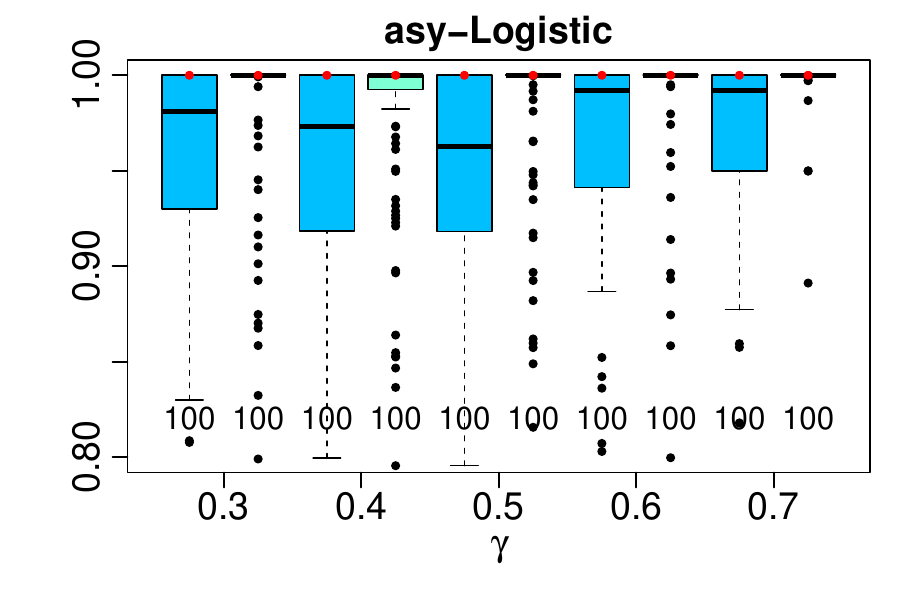}
    \includegraphics[width=0.4\textwidth]{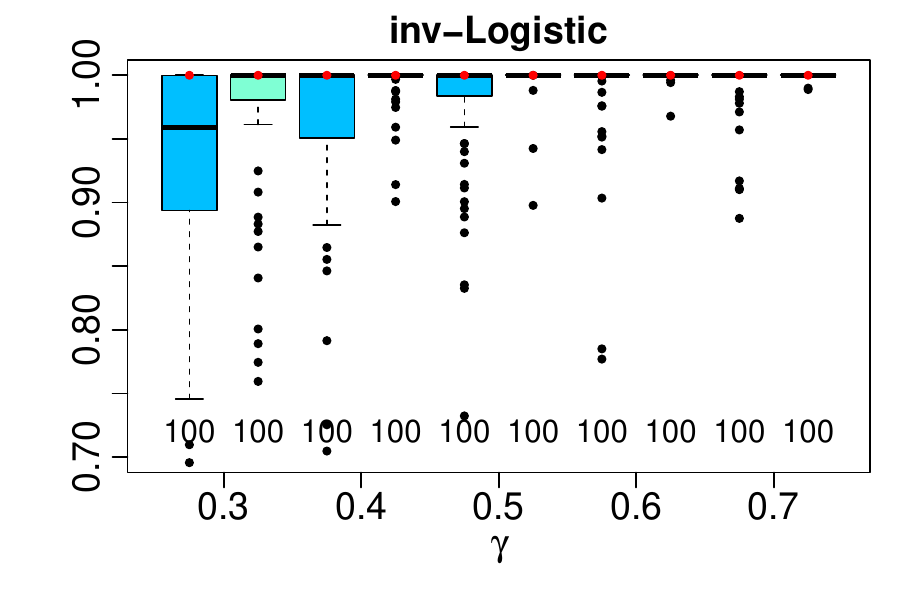}
    \caption{Sampling distribution of the posterior medians of $\tau_1(0.25)$.}
    \label{f:MCMC_tau_600}
    \end{subfigure}
     \caption{Sampling distribution of the posterior medians of $\lambda(0.40)$ and $\tau_1(0.25)$ based on the \Bezier spline (blue) and Simpson-Tawn (green) estimators for four dependence copulas, with dependence parameters set to 0.5. The red lines indicate the true values, and coverage of equi-tailed 95\% intervals are noted below each boxplot.}
    \label{f:MCMC_lambda_and_tau_600}
\end{figure}

Figure \ref{f:MCMC_eta_600} shows boxplots of posterior median of $\eta$ for the four dependence copulas based on the \Bezier spline (blue) and Simpson-Tawn estimators (green). Analytical values of $\eta$ are shown as red dots in each plot, and the coverage of equi-tailed 95\% intervals are noted in plain text below each boxplot. Coverage for the Simpson-Tawn estimator is based on 100 bootstrapped samples from each of the 100 datasets. The \Bezier spline estimator has low bias and nominal coverage for estimates of $\eta$. The Simpson-Tawn estimator has nominal or near-nominal coverage for the AI copulas. when $\gamma = 0.7$; however, It has noticeably lower coverage and higher bias than the \Bezier spline estimator for both AD copulas, and shows a significant decline in coverage as the dependence drops for the asymmetric logistic copula. We note that both estimators have much higher variability in this study when the data size is small. Figure \ref{f:MCMC_lambda_and_tau_600} shows boxplots for the posterior medians of $\lambda(0.40)$ and $\tau_1(0.25)$ based on the two estimators, with the dependence parameters set to 0.5. Both estimators are better at estimating $\lambda(\omega)$ than $\tau_1(\delta)$, and the \Bezier spline estimator tends to have better coverage than the Simpson-Tawn estimator. Additionally, the variability is higher due to the smaller sample sizes. Therefore, we conclude that the \Bezier spline estimator can be used for modeling extremal dependence in small datasets, and performs at least as well as the Simpson-Tawn estimator.

\end{appendix}
\clearpage
\bibliographystyle{asa}
\bibliography{sources}